\documentclass[fleqn,usenatbib]{rasti}

\usepackage{newtxtext,newtxmath}

\usepackage[T1]{fontenc}

\DeclareRobustCommand{\VAN}[3]{#2}
\let\VANthebibliography\thebibliography
\def\thebibliography{\DeclareRobustCommand{\VAN}[3]{##3}\VANthebibliography}

\usepackage{graphicx}	
\usepackage{amsmath}	

\usepackage{epstopdf}
\usepackage{enumerate}
\usepackage{subfigure}

\usepackage{bm}		
\usepackage{booktabs,caption}
\usepackage[T1]{fontenc}
\usepackage{bbding}
\usepackage[flushleft]{threeparttable}

\title[Quasar feedback in the late stage of ellipticals]{On the dominant role of wind in the quasar feedback mode in the late stage evolution of  massive elliptical galaxies}
\author[B. Zhu et al.]
{Bocheng Zhu$^{1,2}$, Feng Yuan$^{1,2}$, Suoqing Ji$^{1}$, Yingjie Peng$^{3,4}$, Luis C. Ho$^{4,3}$ \\
$^{1}$ Key Laboratory for Research in Galaxies and Cosmology, Shanghai Astronomical Observatory, Chinese Academy of Sciences, \\ 80 Nandan Road, Shanghai 200030, People's Republic of China \\
$^{2}$ School of Astronomy and Space Sciences, University of Chinese Academy of Sciences, No. 19A Yuquan Road, Beijing 100049,\\ People's Republic of China \\
$^{3}$ Department of Astronomy, School of Physics, Peking University, 5 Yiheyuan Road, Beijing 100871, People's Republic of China\\
$^{4}$ Kavli Institute for Astronomy and Astrophysics, Peking University, 5 Yiheyuan Road, Beijing 100871, People's Republic of China\\
}

\date{Accepted XXX. Received YYY; in original form ZZZ}

\pubyear{2023}

\begin{document}
\label{firstpage}
\pagerange{\pageref{firstpage}--\pageref{lastpage}}
\maketitle

\begin{abstract}
In this paper we investigate the role of AGN feedback on the late stage evolution of elliptical galaxies by performing high-resolution hydrodynamical simulation in the {\it MACER} framework. By comparing models that take into account different feedback mechanisms, namely AGN and stellar feedback,  we find that AGN feedback is crucial in keeping the black hole in a low accretion state and suppressing the star formation. We then compare the energy from AGN radiation and wind deposited in the galaxy and find that only wind can compensate for the radiative cooling of the gas in the galaxy. Further, we investigate which plays the dominant role, the wind from the cold (quasar) or hot (radio) feedback modes,  by examining the cumulative energy output and impact area to which the wind can heat the interstellar medium and suppress star formation. Our results indicate that first, although AGN spends most of its time in hot (radio) mode, the cumulative energy output is dominated by the outburst of the cold mode. Second, only the impact area of the cold-mode wind is large enough to heat the gas in the halo, while the hot-mode wind is not. Additionally, the cold-mode wind is capable of sweeping up the material from stellar mass loss. These results indicate the dominant role of cold-mode wind. The limitations of our model, including the absence of jet feedback, are discussed.  
\end{abstract}




\section{Introduction}

Observations show that the massive elliptical galaxies { (i.e., $M_{\star}\ga 10^{11}{\rm M_{\odot}}$)} in the contemporary universe have both low star formation rate (SFR) \citep{kauffmann03, baldry04, balogh04, hogg04} and low AGN luminosity \citep{ho08, ho2009}. These observations raise a question in the study of the late stage evolution of massive elliptical galaxies: how the massive elliptical galaxies maintain their quiescent state for both star formation and black hole activity?

Maintaining a quiescent state for both the star formation and black hole accretion in a massive galaxy is not straightforward due to two reasons. First, cosmological inflow from outside the galaxy can potentially trigger star formation. Cosmological simulations predict inflow rates of up to $100\ \mathrm{M_{\odot}}$/yr in massive galaxies at low redshifts \citep{keres05, dekel09, vdv11, nelson13}. However, despite the abundance of gas, the star formation rate in massive galaxies is typically lower than $1~\mathrm{M_{\odot}}$/yr \citep{kauffmann03} and the luminosity of supermassive black holes (SMBHs) also fall between $10^{-6}\sim10^{-4}~L_{\mathrm{Edd}}$\citep{ho08, ho2009}. Additionally, the replenishment of gas from stellar mass losses through wind and supernovae can also trigger star formation if the gas is able to cool efficiently.

It is evident that significant energy inputs are required to prevent the hot gas in the elliptical galaxies from cooling and maintain a quiescent state for both black hole accretion and star formation in massive galaxies \citep[e.g.,][]{peng10, gabor11, zinger20}. These inputs can come from violent processes such as AGN activity or supernovae type Ia (SN Ia), which heat the gas, prevent it from cooling, or eject it to keep the galaxy quenched. However, it has been shown that supernovae alone are not sufficient to maintain a quiescent state. For example, \citet{2018ApJ...866...70L} compared the effects of supernovae and AGN feedback in suppressing star formation in galaxies of different masses through high-resolution galactic-scale simulations. They found that while supernovae can only keep star formation low in low-mass galaxies, AGN feedback plays a dominant role in suppressing star formation in massive galaxies. This conclusion was later confirmed by \citet{wang19}.

AGN feedback is currently believed to be the most important mechanism for the late-stage evolution of massive elliptical galaxies, as supported by both cosmological simulations, such as IllustrisTNG \citep{pillepich18a, zinger20}, and EAGLE \citep{crain15}, and galactic-scale simulations \citep[e.g.,][]{ciotti10, gaspari12, gan14, eisenreich17, yuan18, yoon18, wang20, Su2021}. The observed relationships between the properties of supermassive black holes (SMBHs) and their host galaxies, including the SMBH mass and stellar velocity dispersion, stellar mass, and effective radius \citep{magorrian98, tremaine02, gultekin04}, suggest a co-evolutionary relationship \citep{kormendy13}. Direct observations, such as the presence of bubbles and ripples { at X-ray band}, provide further evidence for AGN feedback \citep{fabian06, wise07, baldi09, blanton11}. { Here ``bubbles'' {are} low density region within the hot gas.} This is reviewed in detail by \citet{fabian12}.

Based on the theory of black hole accretion, the accretion mode can be divided into cold and hot accretion modes bounded by $2\%$ Eddington luminosity \citep{yuan14}. They correspond to two feedback modes, i.e., quasar/cold and radio/hot modes. Different works focus on different modes, although in some early works, the two modes are not explicitly discriminated \citep{hopkins06, matteo05, springel05, Ostriker10, gaspari12, li15, weinberger18, wang19}. Usually, the quasar mode is believed to be responsible for the quenching of galaxies \citep{hopkins06, matteo05, springel05}. The importance of correctly adopting both the cold and hot modes is emphasized by \citet{yoon19}. For example, they show that if only the cold mode were adopted no matter what the accretion rate is, the star formation rate would be two orders of magnitude lower and the fraction of energy ejected in the quasar mode would be too small to be consistent with observations.

The role of AGN feedback on the late-stage evolution of massive galaxies has been explored in some previous works. For example, observed X-ray bubbles as mentioned above are believed to balance the radiative cooling through the work done on the surrounding gas \citep{Churazov05, Rafferty2006, Nulsen2007, HL2012, fabian12}. However, it is unclear whether these bubbles were produced by the current radio mode of the AGN or by a more active quasar mode\footnote{In fact, in Section \ref{gr} of the present paper, we show that the wind in the quasar (cold) feedback mode can well produce a bubble structure.}. On the theoretical front, numerical simulations have shown that kinetic feedback from AGN wind or jet can suppress the cooling flow and maintain a quiescent state in galaxies \citep{gaspari12, li15, choi2015,Su2021}. For example, \citet{Su2021} have investigated the role of jets by performing a series of simulations with different energy forms, including thermal energy, kinetic energy, magnetic energy and cosmic ray (CR), and jet parameters. They found that CR jets appear to be the most promising mechanism for long term quenching. However, these simulations often do not clearly discriminate between the cold and hot modes. Both the quasar mode \citep[e.g.,][]{2017A&A...601A.143F} and the radio mode \citep[e.g.,][]{yuan15, 2021NatAs...5..928S} can produce AGN wind and jet. It is not yet clear which mode plays the dominant role in the late-time evolution of massive galaxies. Although the power of wind in the radio mode is weaker than that in the quasar mode, the black hole spends most of its time in the radio mode \citep[e.g.,][]{yuan18}, so the cumulative effect of wind in the radio mode is potentially significant.

An exception to this trend is the study by \citet{yoon19}, which focuses on the role of the hot mode in AGN feedback. By turning off the hot mode and comparing the simulation results with the fiducial model, they find that star formation rate is hardly changed, suggesting that the hot mode is not important. However, there are two limitations to this study. First, it does not take into account cosmological inflow, which is believed to have an impact on the evolution of the galaxy. Recently, \citet{Zhu2023} investigated the impact and fate of cosmological inflow in massive elliptical galaxies and found that the inflow is blocked at $\sim$ 20 kpc by the pressure gradient from thermalized stellar wind and has little effect on the star formation rate in the whole galaxy. However, for a more accurate conclusion, it would be necessary to revisit the problem after taking into account cosmological inflow. Secondly, and even more importantly, the approach of turning off the hot mode in the simulations may not be very reliable. This is because, once we turn off the hot mode, the gas evolution and properties will adjust correspondingly; thus the cold mode will also inevitably change because in our model the black hole accretion rate is self-consistently determined. Consequently, the evolution of the galaxy will become unrealistic. This call into question the validity of {\citet{Zhu2023}} conclusion. 

Given the limitations of previous studies, in this paper, we aim to determine which mode of AGN feedback, quasar or radio, is dominant in the late-stage evolution of massive elliptical galaxies. Our study will be conducted in {\it MACER} framework, which is dedicated to simulating the evolution of a single galaxy and focuses on the role of AGN feedback. Unlike \citet{yoon19}, we will directly analyze the simulation results of the fiducial model, which properly takes into account both the cold and hot modes, to compare the roles of the two modes. The paper is structured as follows. In Section \ref{model}, we review the basic aspects of {\it MACER}. The simulation setup is introduced in Section \ref{setup}. Our main results are presented in Section \ref{res}. Finally, we summarize and discuss our results in Section \ref{conclusion}.

\section{The {\it MACER} Model}\label{model}

The simulations performed in this paper use the suite of the {\it MACER} project, which is developed based on early works \citep{ciotti01,ciotti09,ciotti10, Ostriker10,novak11,gan14}. It is a two-dimensional radiation hydrodynamical simulation based upon ZEUSMP/2 code, simulating the evolution of a single galaxy. The radiative transfer is included in the simulation by solving the simple one-dimensional radiative transfer equation in the radial direction and ignoring the scatter. The most updated version is presented in \citet{yuan18}, in which the state-of-the-art AGN physics, including radiation and wind as a function of accretion rates in both the hot and cold accretion (feedback) modes, has been incorporated. Most recently, \citet{Zhu2023} have included the stellar yields, AGN feedback at super-Eddington regime and external gas supply from cosmological inflow.

The inner boundary of the simulation is chosen to be smaller than the Bondi radius. In this case, we can combine the calculated mass flux at the inner boundary with the black hole accretion theory to obtain a reliable value of the accretion rate at the black hole horizon. This accretion rate determines the power of the AGN and thus is a crucial parameter in the study of AGN feedback. The highest resolution is achieved at the inner region where the interaction between AGN outputs and the gas in the galaxy is the strongest, which is as high as 0.5 pc in the case of a massive elliptical galaxy. However, current version of {\it MACER} is lacking jet feedback and dust, we will include them in our future update. Below we introduce several other key features of {\it MACER}.

\subsection{AGN physics}
\label{AGNphysics}

The black hole accretion process in {\it MACER} is divided into two modes: the cold mode and the hot mode, depending on the value of the AGN luminosity or black hole accretion rate. When the mass accretion rate is greater than a critical mass accretion rate, the accretion mode will change from cold mode to hot mode and vice versa. The critical mass accretion rate is $\sim 2\%\dot{M}_{\rm Edd}$ based on the observation of X-ray binaries \citep{mc06}. The AGN physics in {\it MACER} incorporates the most recent developments in the black hole accretion theory for both modes, including the wind in both the cold and hot modes and the radiation in the hot mode as a function of the black hole accretion rate.

In the cold mode, we assume that the inflow through the inner boundary of our simulation domain forms an accretion disk at the circularization radius, and the black hole accretion rate is calculated by solving a set of ordinary differential equations (ODEs) that describe the mass evolution of the accretion disk and the mass loss in the wind. The set of ODEs is described as follows:

\begin{equation}
\frac{d\dot{M}_{\rm eff}}{dt} = \frac{\dot{M}(r_{\rm in})-\dot{M}_{\rm eff}}{\tau_{\rm ff}},
\end{equation}
\begin{equation}
M_{\rm dg} = \int \dot{M}_{\rm eff} dt,
\end{equation}
\begin{equation}
\dot{M}_{\rm d, inflow} = \frac{M_{\rm dg}}{\tau_{\rm vis}},
\end{equation}
\begin{equation}
\dot{M}_{\rm BH} = \dot{M}_{\rm d, inflow}-\dot{M}_{\rm wind}.
\end{equation}
In these ODEs, $\dot{M}(r_{\rm in})$ is the inflow rate at the inner boundary, $\dot{M}_{\rm eff}$ is the effective accretion rate that the gas falls into the small disk, $\tau_{\rm ff}\equiv r_{\rm in}/(2GM_{\rm BH}/r_{\rm in})^{1/2}$ is the free-fall time scale from the inner boundary to the small disk, $M_{\rm dg}$ is the total mass of small disk, $\tau_{\rm vis}\equiv 1.2\times10^6~(M_{\rm BH}/10^9~{\rm M_{\odot}})~ \mathrm{yr}$ is the instantaneous viscous timescale, $\dot{M}_{\rm d, inflow}$ is the accretion rate from the small disk to the accretion disk, and $\dot{M}_{\rm wind}$ is the mass loss rate via the disk wind.

The calculated accretion rate is used to further divide the cold mode into the standard thin disk and super-Eddington regimes, { separated} by $\dot{M}_{\rm Edd}$. It is important to note that there is no Eddington limit for the emitted AGN luminosity, as indicated by both accretion theory \citep[e.g.,][]{Abramowicz1988,2014MNRAS.439..503S, 2014ApJ...796..106J} and observations \citep[e.g.,][]{2013ApJ...764...45K}.

In the regime of the standard thin disk, the radiative efficiency of accretion flow is set to 0.1. The mass flux and velocity of wind are taken from the statistical results obtained by analyzing a sample of luminous AGNs \citep{gofford15}. They can be well described as a function of bolometric AGN luminosity $L_{\mathrm{bol}}$:
\begin{equation}
    \dot{M}_{\mathrm{wind,cold}} = 0.28\left(\frac{L_{\mathrm{bol}}}{10^{45} \mathrm{erg\ s^{-1}}}\right)^{0.85}\ \mathrm{M_{\odot}\ yr^{-1}}
\end{equation}
\begin{equation}
    v_{\mathrm{wind,cold}} = \max\left(2.5\times 10^4 \left(\frac{L_{\mathrm{bol}}}{10^{45} \mathrm{erg\ s^{-1}}}\right)^{0.4}, 10^5\right)\ \mathrm{km\ s^{-1}}
\end{equation}

In the super-Eddington regime, there is limited observational data. Recent work by \citet{2022arXiv221110710Y} has analyzed three-dimensional radiative GRMHD numerical simulation data of super-Eddington accretion around a massive black hole and obtained the mass flux and velocity of wind as a function of accretion rate. These results are adopted in the {\it MACER} model.  The radiative efficiency of super-Eddington accretion flow is obtained by fitting the three-dimensional radiative GRMHD simulation results of \cite{jiang19}. The results are:
\begin{equation}
    \dot{M}_{\mathrm{wind,super}} = \left(\frac{r_{\rm d}}{45 r_s}\right)^{0.83}\dot{M}_{\rm BH},
\end{equation}
\begin{equation}
  v_{\mathrm{wind,super}} = 0.15~c,  
\end{equation}
\begin{equation}
    \varepsilon_{\mathrm{super}} = 0.21\left(\frac{100\dot{M}_{\mathrm{BH}}}{\dot{M}_{\mathrm{Edd}}}\right)^{-0.17}.
\end{equation}
where $r_{\rm d}$ is the outer boundary of the super-Eddington accretion flow, which is determined by the specific angular momentum of the inflowing gas in our simulation. The angle distribution of the mass flux is set to $0^{\circ}-30^{\circ}$ and $150^{\circ}-180^{\circ}$.

In the hot mode, the geometry of the accretion flow is described by an inner hot accretion flow and an outer truncated thin disk and the boundary between them is determined by the truncated radius $r_{\rm tr}$, which is a function of the accretion rate \citep{yuan14}{:
\begin{equation}
    r_{\mathrm{tr}} \approx 3 R_{\mathrm{s}}\left[\frac{2\times10^{-2}\dot{M}_{\mathrm{Edd}}}{\dot{M}(r_{\rm in})}\right]^2\label{con:rtr},
\end{equation}
where $R_{\mathrm{s}}\equiv 2G M_{\rm BH}/c^2$ is the Schwarzschild radius. }

Strong wind is observed in numerical simulations of hot accretion flows \citep{2012ApJ...761..130Y,yuan15,yang21} and observations \citep{2013Sci...341..981W,2016Natur.533..504C,2019MNRAS.483.5614M,2019ApJ...871..257P,2021NatAs...5..928S}. Since observational constraints on wind properties are still weak, we adopt the mass flux and velocity of the hot wind from the theoretical results in \citet{yuan15}. The wind velocity and mass flux can be described by:
\begin{equation}
    v_{\mathrm{wind,hot}} = 0.2 v_{\rm k}(r_{\rm tr}),
\end{equation}
\begin{equation}
    \dot{M}_{\mathrm{wind,hot}} = \dot{M}_{{\rm in}}\left(1-\left(\frac{3r_{\rm s}}{r_{\rm tr}}\right)^{0.5}\right),
\end{equation}
where $\dot{M}_{{\rm in}}$ is the gas flow rate at the inner boundary of the simulation region { and $v_{\rm k}(r_{\rm tr}) = \sqrt{G M_{\rm BH}/r_{\rm tr}}$ is the Keplerian velocity at the truncated radius $r_{\rm tr}$.}  { The radiation efficiency $\varepsilon_{\mathrm{hot}}$ of a hot accretion flow is described as
\begin{equation}
   \varepsilon_{\mathrm{hot}} (\dot{M}_{\mathrm{BH}})=\varepsilon_0\left(\frac{\dot{M}_{\mathrm{BH}}}{0.1L_{\mathrm{Edd}}/c^2}\right)^a.\label{con:eps}
\end{equation}
The values of parameter $\varepsilon_0$ and $a$ can be found in \cite{xie12}.} 

In the calculation of AGN radiation, we consider both radiative heating and radiation pressure. The key parameter for determining radiative heating is the Compton temperature, which is taken from the works of \citet{sazonov05} and \citet{xie17} for the cold and hot accretion modes, respectively. { The Compton temperature in the cold mode is set to $2\times10^7{\rm K}$, while for the hot mode the Compton temperature $T_{\rm comp}$ is set to} 
\begin{equation}
T_{\rm comp}=
    \begin{cases}
    4\times10^7{\rm K}&10^{-3}\la L_{\rm BH}/L_{\rm Edd}\la 0.02\\
    10^8 {\rm K}&L_{\rm BH}/L_{\rm Edd}\ga 10^{-3}.
    \end{cases}
    \label{con:Tcomp}
\end{equation}

Regarding the AGN wind, we incorporate its energy, momentum, and mass flux as a source term in the innermost two grids. This wind will then propagate to larger scales and interact with the  gas within the galaxy. These interactions are automatically calculated by the code.

\subsection{Star formation and stellar feedback model}

The code {\it MACER} also considers star formation, stellar mass loss, stellar yields, and stellar feedback. The star formation rate is calculated using the following equation:
\begin{equation}\label{sfr_formula}
\dot{\rho}_{\mathrm{SFR}} = \frac{\eta_{\mathrm{SF}}\rho}{\tau_{\mathrm{SF}}},
\end{equation}
where $\eta_{\mathrm{SF}}$ is the star formation efficiency, which is set to 1\%, and $\tau_{\mathrm{SF}}$ is the star formation time scale, as described in \citet{yuan18}.

Given the limited resolution of the code, star formation is considered to occur when the gas number density is higher than $1~\mathrm{cm^{-3}}$ and the temperature is lower than $4\times 10^4~\mathrm{K}$, as is commonly adopted in simulation works.

The stellar feedback model in {\it MACER} includes the thermalization of mass loss from old stars, feedback from Type Ia supernovae (SN Ia), and feedback from Type II supernovae (SN II). The calculations of the stellar mass-loss rate and stellar yields from old stars are based on stellar evolution theory \citep{maraston05}.

In our simulation, we assume that the simulated galaxy formed at high redshift { (z $\ga$ 5)} through an intense star formation {burst} and has a uniform age of the stellar population. The age of the stellar population is set to 2 Gyr at the start of the simulation { and our simulations last for 12 Gyr.} The radial profile of the metallicity of stars is based on observations \citep{spolaor11}, and can be described as:
\begin{equation}
    (Z/Z_{\odot})(r) = \exp[-0.23\log(r/r_{\rm e})]+0.3.
\end{equation}

The feedback from SN Ia is modeled as the ejection of $m_{\mathrm{Ia}}=1.4~\mathrm{M_{\odot}}$ of mass and $E_{\mathrm{Ia}}=10^{51}~{\rm erg}$ of energy per SN Ia event. The energy released by SN Ia is deposited locally in the form of thermal energy. 

The limited resolution might not properly capture evolution of the thermalized SN remnant because when the resolution is poor, the energy will be distributed to too large space (thus too much mass). In this case, the radiative cooling timescale will become artificially long therefore the sound crossing timescale may be longer than the cooling time scale; consequently, the injected energy will be quickly radiated away. However, since our simulated galaxy is massive elliptical galaxies, most of the gas is hot and tenuous gas thus the cooling timescale is usually long and such a problem will be alleviated. But even in this case, the thermal injection does weaken the SN Ia feedback in some some degree in our simulations. Recent works \citep[e.g.,][]{li20a, li20b} investigate this issue in hot medium with high resolution local box simulations. They find that the effective cooling rate will be suppressed due to the inhomogeneity of the medium. To compensate this effect, we have increased the injected thermal energy of the stellar feedback by a factor of 1.4 (Miao Li, private communication).

The mass return from massive stars is calculated as the IMF-weighted average of the mass loss of stars with masses greater than 8 solar masses, with the upper limit of the IMF set to $40~\mathrm{M_{\odot}}$.

\subsection{Cosmological inflow in {\it MACER}}

\citet{Zhu2023} have investigated the impacts of the cosmological inflow and found that it does not have much impacts on the late-stage evolution of massive elliptical galaxies. But for completeness, in the present work we include  cosmological inflow. 
Specifically, the information of cosmological inflow comes from the public data of TNG100, which is part of the IllustrisTNG cosmological simulation project\footnote{https://www.tng-project.org/data/} \citep{springel18, pillepich18b, naiman18, nelson18, marinacci18}, for the properties of inflow, such as the mass inflow rate.

In our simulations, for technical convenience, we do not exactly use the results obtained above. Instead, we simply take them as our important reference and adopt the following inflow properties: 1) the gas inflow rate is set to $100~\mathrm{M_{\odot}}$/yr; 2) the radial velocity is set to 0.5 times the virial velocity; 3) the square of the sound speed of the inflow is set to 0.8 times the square of the virial velocity; 4) the metallicity is set to 0.18 times the solar metallicity; 5) the rotational velocity is set to 0. The geometry of the inflow is spherical. This setup of geometry is also consistent with previous works \citep{keres05, nelson13} {that indicates that infall becomes more spherical at lower redshift.} We assume that the inflow properties do not change over time and that only hot spherical inflow is taken into account, as cold inflow is not considered to be important in low-redshift massive galaxies. 


\subsection{Galaxy model}
Following \cite{yuan18}, the dark matter halo and stars of our elliptical galaxy have a static, spherically symmetric distribution. The initial total stellar mass $M_{\star}$ is set to $3\times 10^{11}~\mathrm{M_{\odot}}$. The stellar distribution is adopted to the Jaffe profile \citep{jaffe83}:
\begin{equation}
    \rho_{\star} = \frac{M_{\star} r_{\rm J}}{4\pi r^2 (r_{\rm J}+r)^2},
\end{equation}
where $r_{\rm J}$ is the Jaffe radius and is set to $9.04\ \mathrm{kpc}$ in our simulations. The corresponding effective radius $R_{\rm e}$ is $6.9\ \mathrm{kpc}$. The one dimensional stellar velocity dispersion $\sigma_0$ is set to $260\ \mathrm{km/s}$. The effective radius $R_{\rm e}$ and the stellar velocity dispersion $\sigma_0$ are chosen to let the galaxy obey the fundamental plane \citep{djorgovski87} and Faber-Jackson relation \citep{faber76}. The black hole mass $M_{\mathrm{BH}}$ is set to $1.8\times 10^9~\mathrm{M_{\odot}}$, which obey the $M_{\mathrm{BH}}-\sigma$ relation \citep{kormendy13}.

Observations shows that the total mass density distribution is well described by $r^{-2}$ for a large radial range \citep{dye08}. In this work, we use the same setup as the previous work in this series. The total density profile is assumed to be an isothermal sphere:
\begin{equation}
\rho_{\mathrm{total}} = \frac{v_{\mathrm{c}}^2}{4\pi G r^2},
\end{equation}
where $v_{\mathrm{c}}=\sqrt{2}\sigma_0$. The dark matter mass and virial radius are set to $2\times10^{13}~\mathrm{M_{\odot}}$ and 513 kpc, respectively, as in previous works in this series.

{ We assume that the stellar wind will be thermalized due to their collision. The sound speed  will roughly be equal to the local velocity dispersion of stars.} To calculate the thermalized temperature of the stellar wind, we solve the Jeans equation to determine the stellar velocity dispersion, $\sigma_{\star}$:
\begin{equation}
\frac{dP}{dr} = -\rho_{\star}\nabla (\Phi_{\mathrm{BH}}+\Phi_{\mathrm{Iso}}),
\end{equation}
where $P = \rho_{\star}\sigma_{\star}^2$.

\begin{figure}
    \includegraphics[width=0.45\textwidth]{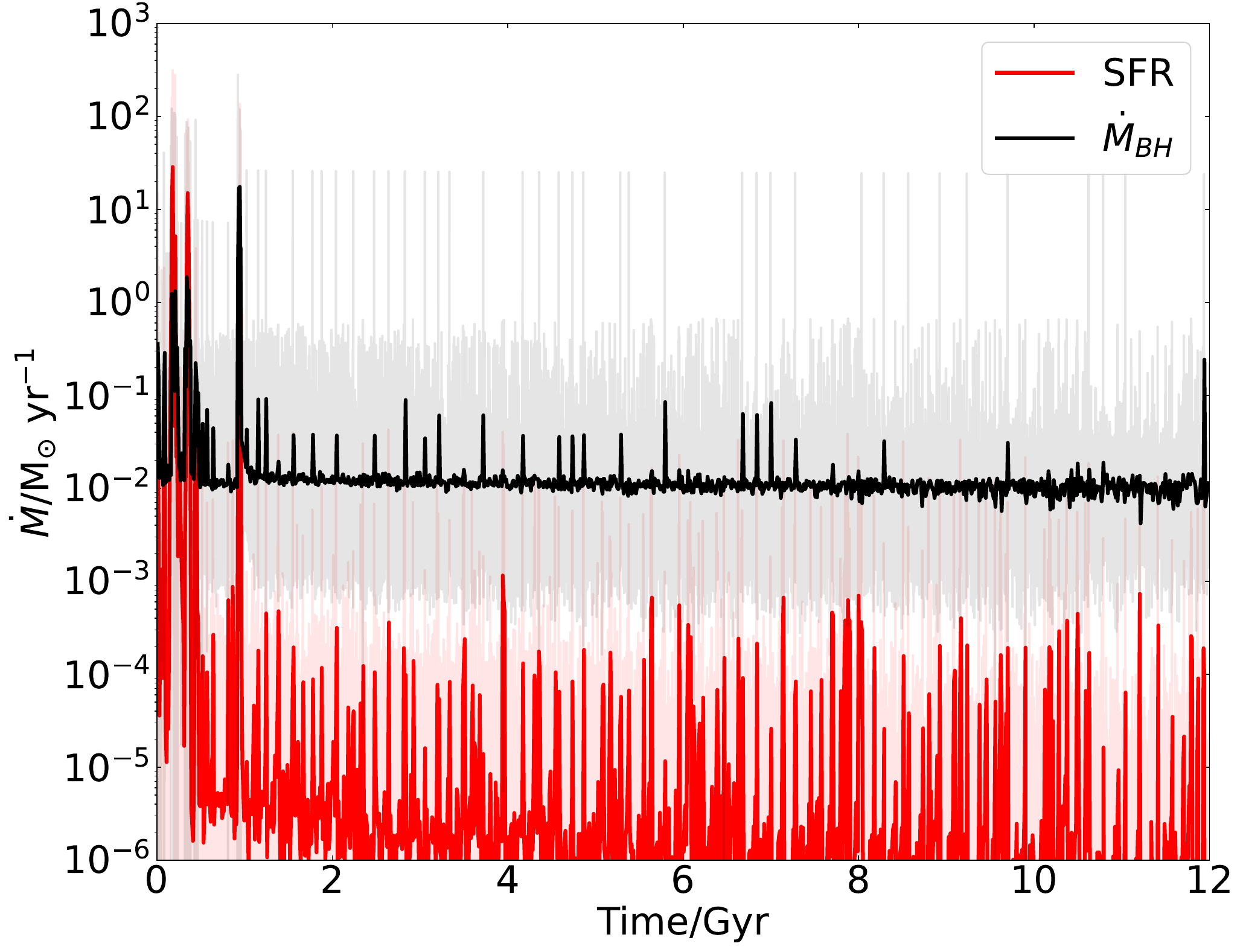}
    \caption{Time evolution of SFR and BHAR in the ``Fiducial'' model. The translucent lines represent data with a time interval of $10^4$ yr, while the opaque lines represent data with a 100 Myr average.}
    \label{SFR&BHAR}
   \end{figure}

\begin{figure*}     
    \includegraphics[width=\textwidth]{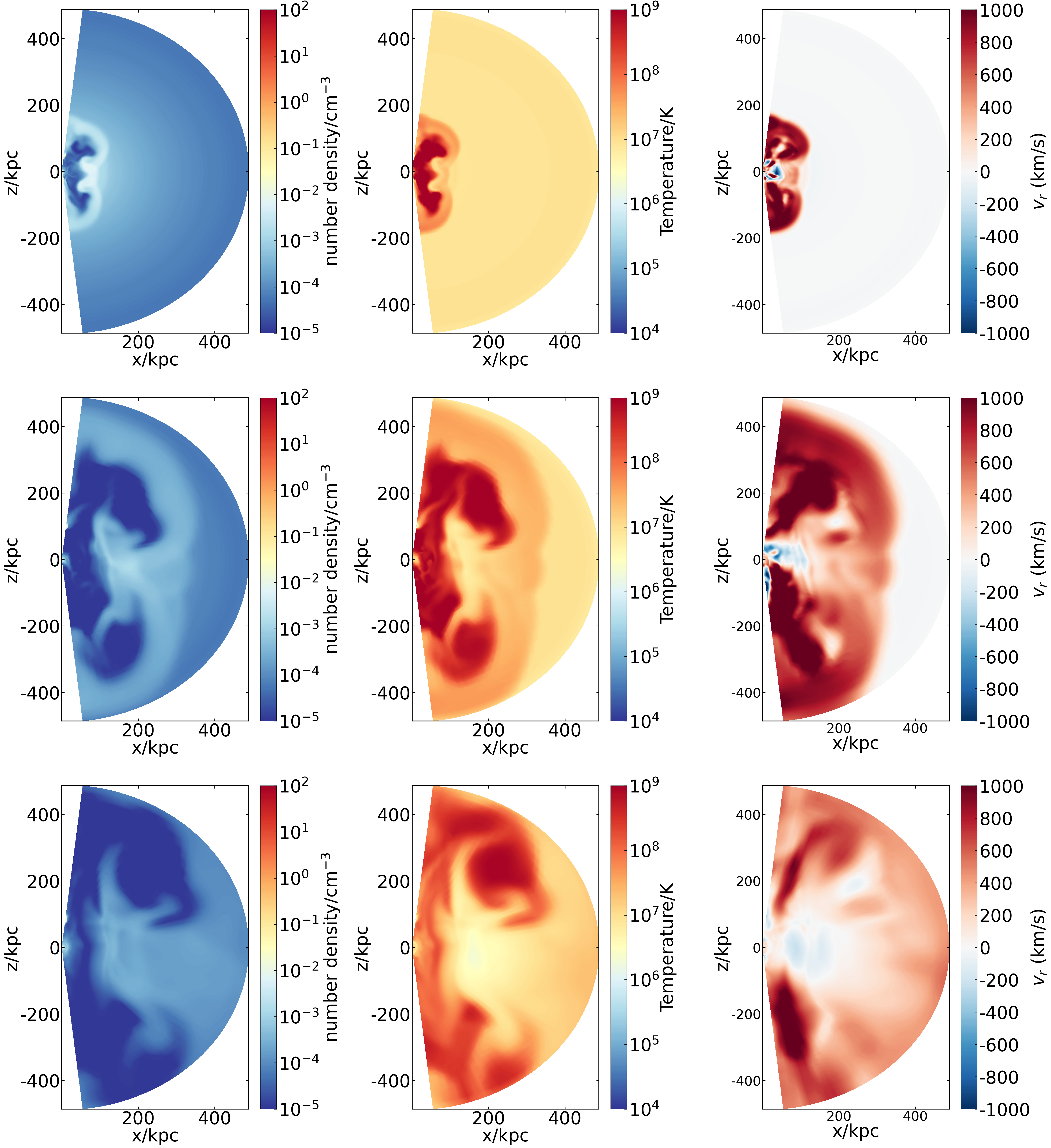}
    \caption{The spatial distribution of the gas density, temperature, and radial velocity at 0.25 (top row), 0.5 (middle row), and 0.75 Gyr (bottom row).}
    \label{snap_early}
\end{figure*}   
   
\section{Simulations}\label{setup}

In this paper, we analyze the simulation data from three models presented in \citet{Zhu2023}: the ``Fiducial'', ``NoAGN'', and ``NoSN'' models. The ``Fiducial'' model includes all forms of feedback, while the NoAGN'' and ``NoSN'' models are obtained by turning off AGN and SN feedback, respectively, in the ``Fiducial'' model at $t=6$ Gyr.

The simulations are carried out under two-dimensional axisymmetric spherical coordinates ($r,\theta,\phi$), which allows for the consideration of the rotation of the flow. The inner and outer boundaries of the simulation domain are located at 2.5 pc and 500 kpc, respectively. The inner boundary is typically ten times smaller than the Bondi radius for elliptical galaxies. The simulation has a resolution of $240\times 60$ in the $r-\theta$ plane, with a homogeneous mesh in the $\theta$ direction and a logarithmically spaced mesh in the radial direction. The inner and outer radial boundaries are set to be inflow-outflow boundaries, while the boundary at the pole is set to be a symmetry boundary.

{ The stellar mass and halo mass for our simulated galaxies is $3\times10^{11}{\rm M_{\odot}}$ and $2\times10^{13}{\rm M_{\odot}}$, respectively.} The initial gas profile in our simulations is described by a $\beta$ profile: 
\begin{equation}
    \rho_{\mathrm{gas}}(r)=
    \begin{cases}
        \rho_0 (1+(r/r_{\rm c})^2)^{-3/2\beta} & r> r_{\rm c}\\
        \rho_0/2\ (r/r_{\rm c})^{\alpha} & r\le r_{\rm c}\\ 
    \end{cases}\label{con:dens}
\end{equation}
This profile is characterized by parameters $r_{\rm c}$, which is set to 6.9 kpc, $\beta$, which is set to $2/3$ based on observational data \citep{anderson13}, the power-law index $\alpha$, which is set to $-1$, and $\rho_0$, which is set to $2.5\times10^{-1} { {\rm cm^{-3}}}$. The corresponding baryon fraction is $\sim50\%$ cosmic mean baryon fraction. The gaseous halo is assumed to be hydrostatic, and its gas pressure distribution is calculated by integrating the hydrostatic equation.
Note that our setup for the gas profile has larger $r_{\rm c}$ than observations of NGC 4472 \citep{werner2012}, resulting in an extended core. Since the gas within the extended core will be more vulnerable to thermal instability due to their high density, the star formation rate and black hole accretion rate in our simulations are significantly higher than that typically observed in elliptical galaxies at the beginning. As we will discuss in section \ref{gr}, due to the initial high black hole accretion rate, a strong AGN outburst will occur and drive the galaxy into quiescent state.


 \section{Results}\label{res}

\subsection{Overview of the galaxy evolution}
\label{gr}  

Figure \ref{SFR&BHAR} illustrates the evolution of the star formation rate (SFR) and black hole accretion rate (BHAR) in the ``Fiducial'' model of our simulation. Translucent lines represent data at $10^4$~year intervals, while opaque lines denote 100 Myr averages. During the first 1 Gyr, both the SFR and BHAR are elevated, reaching values above $10~{\rm M_{\odot}/yr}$. This high SFR and BHAR can be attributed to the high-density extended core in our initial conditions. { The gas within the extended core will be more vulnerable for thermal instability due to their high density}. In addition, there is another important reason for the high SFR, as we will discuss later in this section.

At approximately 1 Gyr, the galaxy undergoes a noticeable ``transition'', as evidenced by the sharp decline in both SFR and BHAR values. Subsequently, these values remain low, with the peak SFR value dropping below $10^{-3}~\mathrm{M_{\odot}/yr}$ and the corresponding specific star formation rate around $10^{-14}~\mathrm{yr^{-1}}$, indicating that the galaxy becomes quenched.

\begin{figure}
    \includegraphics[width=0.48\textwidth]{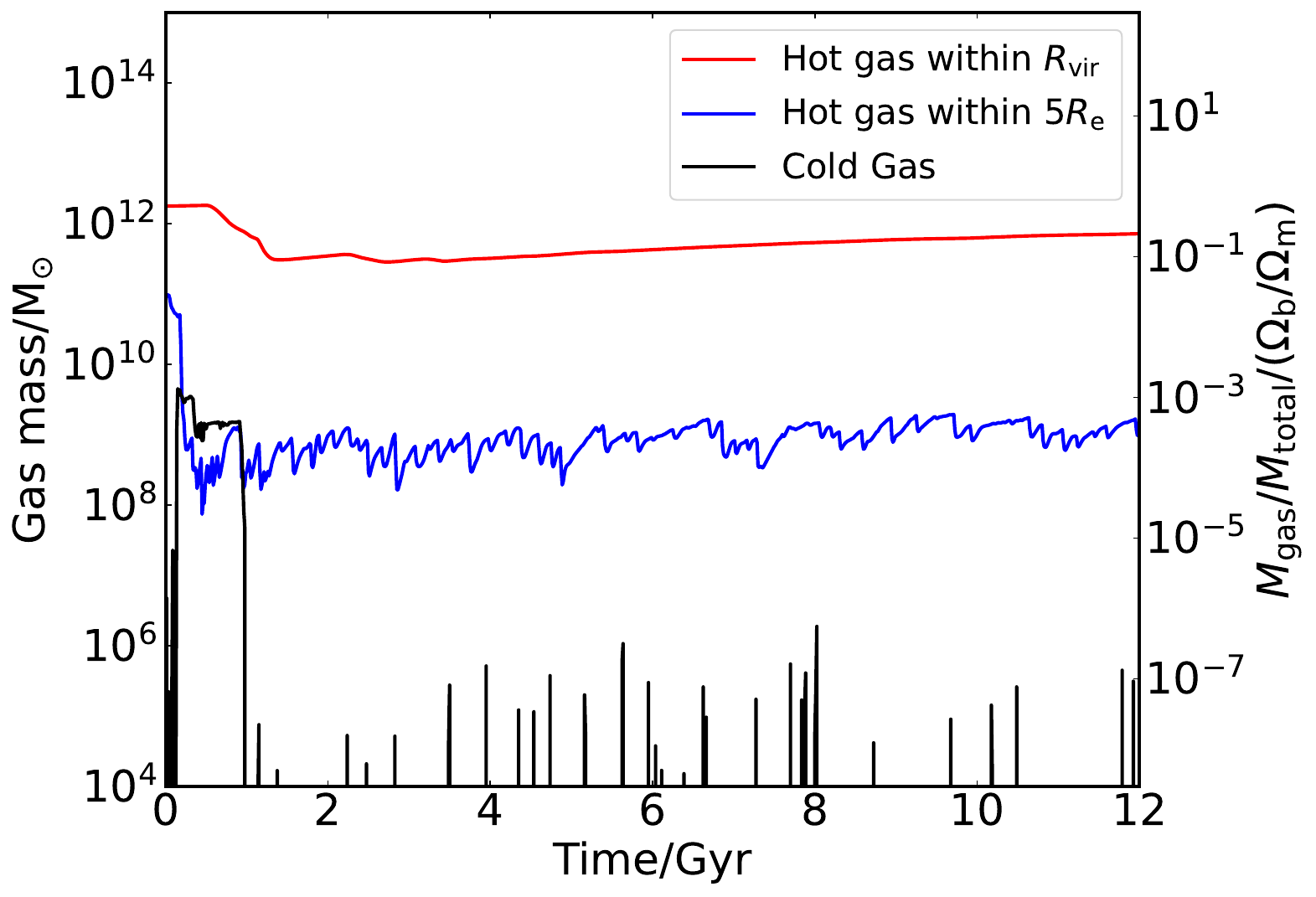}
    \caption{Time evolution of the gas mass budget. The red and blue lines represent the total mass of hot gas within $R_{\rm vir}$ and 5$R_{e}$, respectively;  the black line represents the total mass of cold gas within 5$R_{e}$. The cold gas is defined as the gas with a temperature less than $4\times10^4$ K. The data-dump time interval is 12.5~Myr.}
    \label{gasbudget}
   \end{figure}

\begin{figure*}     
    \includegraphics[width=\textwidth]{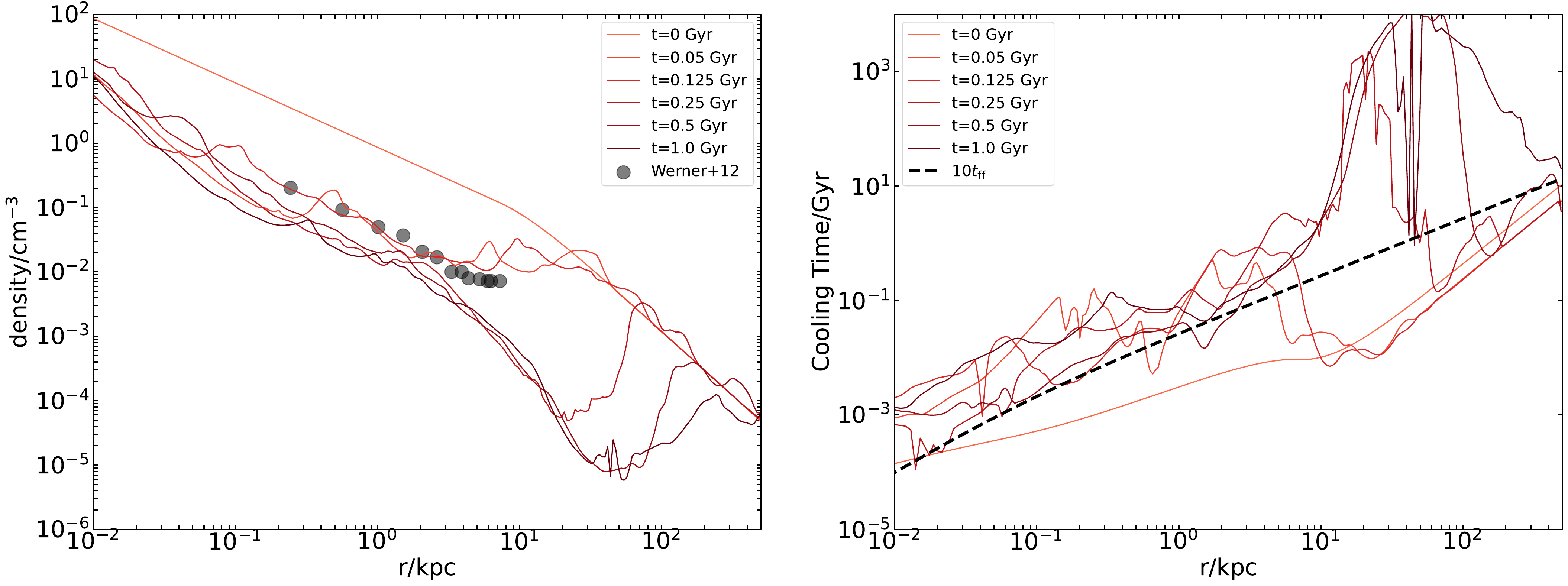}
    \caption{ The time evolution of radial profiles of density (left panel) and cooling time (right panel) of the ``Fiducial'' model during the initial burst. The red lines from light to deep represents the radial profile at 0, 0.05, 0.125, 0.25, 0.5, and 1.0 Gyr. The black filled circles in the left panel represent {the observations of a massive elliptical galaxy NGC 4472 with stellar mass $M_{\star}\sim 2\times10^{11}\mathrm{M_{\odot}}$} from \citet{werner2012}. The black dashed line in the right panel represents the profile of 10 times free-fall time.}
    \label{profile_early}
\end{figure*}
   
\begin{figure*}     
    \includegraphics[width=\textwidth]{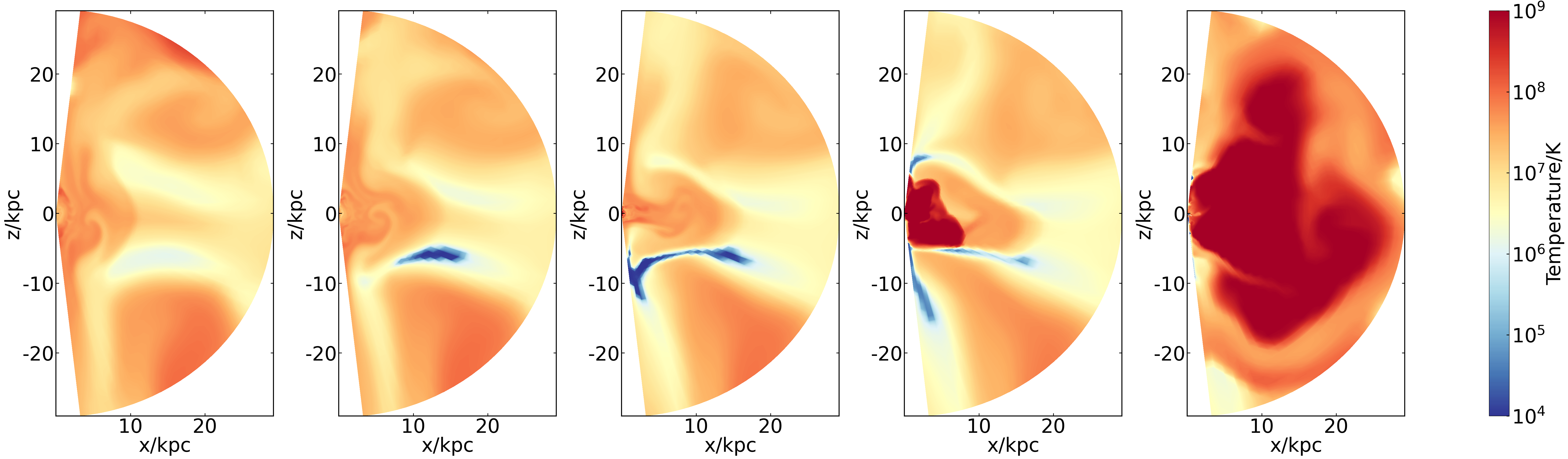}
    \caption{Snapshots of the spatial distribution of the temperature at 125 Myr, 137.5 Myr, 150 Myr, 162.5 Myr, and 175 Myr. These times correspond to the time of the first peak of the cold gas mass evolution curve shown in Figure \ref{gasbudget}.}
    \label{snap_cold}
\end{figure*}

\begin{figure*}     
    \includegraphics[width=\textwidth]{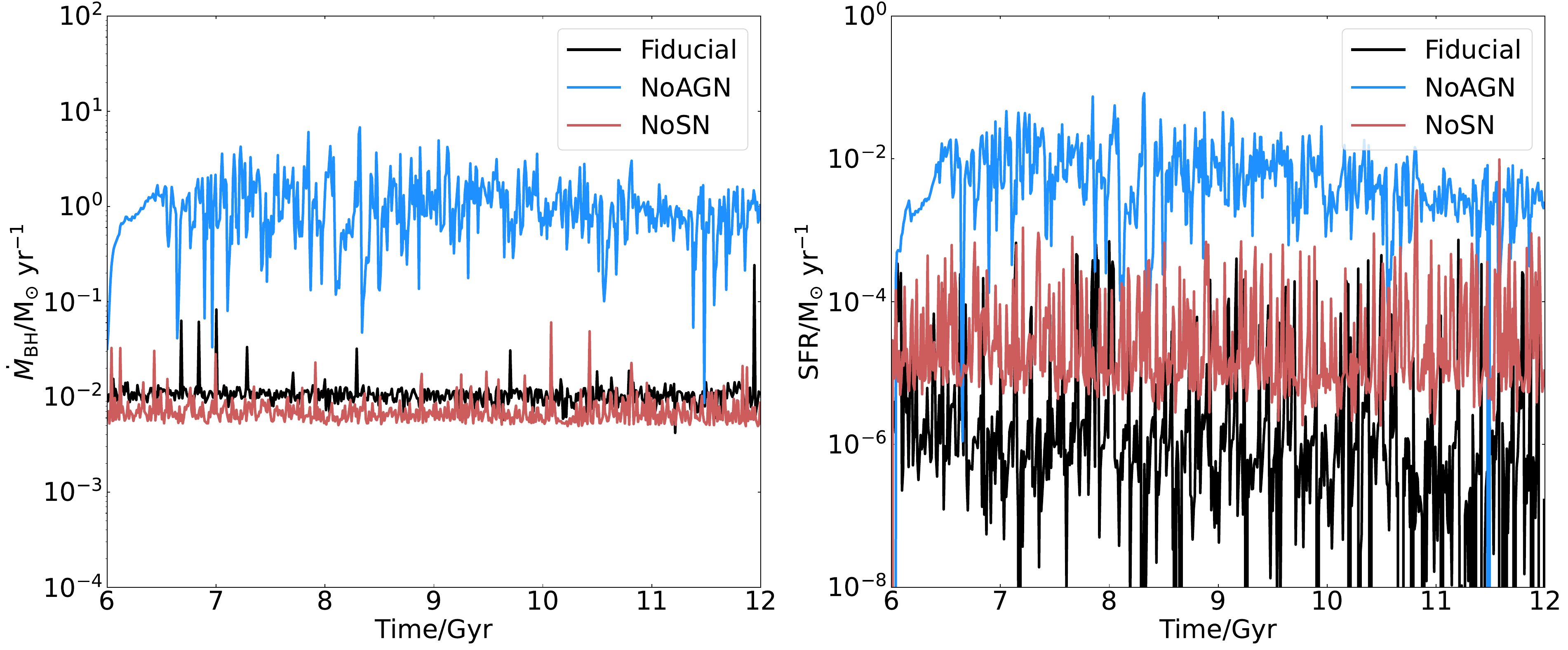}
    \caption{Time evolution of BHAR (left panel) and SFR (right panel) of the ``Fiducial'', ``NoAGN'', and {``NoSN''} models from 6 to 12 Gyr. The data is averaged with 100 Myr.}
    \label{SFR&BHAR_total}
\end{figure*}

What causes this ``transition'' in SFR and BHAR? Figure \ref{snap_early} presents the spatial distribution of gas density, temperature, and radial velocity within the galaxy at 0.25 Gyr (top row), 0.5 Gyr (middle row), and 0.75 Gyr (bottom row), respectively. At 0.25 Gyr, a high-temperature, low density, and rapidly expanding ``bubble'' emerges in the central region of the galaxy. This bubble appears early in the simulation and gradually expands. The bubble's radial velocity exhibits a bipolar structure with a wide opening angle. Given the large BHAR values during this period and the bipolar, wide-angle nature of the AGN wind in our model, this structure is likely a result of the powerful AGN wind's inflation. The temperature within the bubble can sometimes reach $\sim 10^9$ K, which is due to the thermalization of the high-speed AGN wind in our model. 

As time progresses from the top to middle and bottom rows, the bubble continues to expand. At smaller radii, the bubble's bipolar shape is maintained due to the AGN wind's bipolar structure. This suggests that the bubble can only displace gas within the bipolar region. As the wind propagates, its kinetic energy gradually converts into thermal energy through heating caused by the shock resulting from the wind-medium interaction. The figure reveals that the bubble comprises two ``layers'': the inner layer contains shocked AGN wind, while the outer layer primarily consists of a shocked medium around the BH. Since the thermal energy of the gas dominates over the kinetic energy in the bubble, the bubble loses its bipolar shape as it moves outward and expands laterally toward the equatorial region. Consequently, the bubble pushes gas outwards, encompassing the entire galactic region.

At $t\sim1$ Gyr, this bubble moves beyond the outer boundary of our simulation domain. From 1 Gyr until the end of our simulation, new bubbles occasionally appear, but they are significantly less powerful compared to the bubble depicted in Figure \ref{snap_early}. These bubbles do not grow substantially and are typically around 1 kpc in size. This is due to the significant decrease in gas density within the galaxy after 1 Gyr, resulting in a much lower BHAR, consistently below $10^{-1}~{\rm M_{\odot}/yr}$, and a weaker AGN wind.

The continuous inflation of the large bubble shown in Figure \ref{snap_early} leads to a rapid decrease in the total gas content within the galaxy. Figure \ref{gasbudget} displays the time evolution of hot and cold phase gas mass, including hot gas mass within the virial radius $R_{\mathrm{vir}}$ (red line), hot gas within 5 times the effective radius $R_{\rm e}$ (blue line), and cold gas within $5~R_{\rm e}$ (black line). A substantial amount of cold gas forms during the first 1 Gyr, with the peak cold gas mass reaching $\sim 10^{9}~\mathrm{M_{\odot}}$ during the burst phase. This cold gas is likely to significantly enhance both black hole accretion and star formation. In fact, our examination of black hole accretion reveals that approximately 90\% of the mass accretion rate is supplied by cold gas. The hot gas mass within $R_{\mathrm{vir}}$ and $5~R_{\rm e}$ declines significantly within the first 1 Gyr. The amount of hot gas within $5~R_{\rm e}$ drops from $\sim 10^{11}~\mathrm{M_{\odot}}$ to $\sim 10^{9}~\mathrm{M_{\odot}}$, while the total mass of hot gas within $R_{\mathrm{vir}}$ decreases from $\sim10^{12}~\mathrm{M_{\odot}}$ to $\sim 10^{11}~\mathrm{M_{\odot}}$. The former occurs within 0.5 Gyr, while the latter takes place over a longer timescale of $\sim$ 1 Gyr. The different timescales reflect the time required for the bubble to propagate from $5~R_{\rm e}$ to $R_{\mathrm{vir}}$, as the AGN wind originates from the center and moves outward. This suggests that the large bubble formed during the initial $\sim$ 1 Gyr can reach the virial radius and push gas out of the halo.

{ We have further investigated the time evolution of the radial profile of the gas density and cooling time during the initial burst. The results are shown in Figure \ref{profile_early}.  From the figure, we can see that the gas density within 10 kpc decreases by nearly 2 order of magnitude in the first 50 Myr, then such a low-density region expands gradually from 10 kpc to $>100$ kpc from 50 Myr to 1 Gyr. This result is due to the inflation of the bubble shown in Figure \ref{snap_early}. After the initial burst, the value of gas density is roughly consistent with the observations.

For the evolution of the cooling time, we can see from the figure that the cooling time of the gas within 10 kpc at the beginning is less than $\sim$ 10 Myr, implying that the cooling catastrophe will happen $\sim$10 Myr after the simulation begins. Also shown in the figure is 10 times of free fall time $10 t_{\rm ff}$. The ratio of cooling time and $10 t_{\rm ff}$ can be used to judge the thermal stability of the gas halo \citep{sharma12}, namely the local thermal instability will be triggered when the cooling time is shorter than $10 t_{\rm ff}$. From the figure, we can see that this criterion is satisfied at the beginning; but after the initial burst, the median cooling time is slightly close to $10 t_{\rm ff}$, implying that the initial burst converts the gaseous halo from thermally unstable to moderately thermally stable.}

We now explore the factors contributing to the high BHAR during the first 1 Gyr. One reason is the high gas density { in the extended core} in our initial conditions. Another factor is that the bubble depicted in Figure \ref{snap_early} strongly perturbs and compresses the ambient gas around its edge as it propagates outward, leading to the formation of cold gas at the bubble's boundary. This is demonstrated in Figure \ref{snap_cold} and is also responsible for the sharp increase in cold gas mass at $\sim$ 1 Gyr shown in Figure \ref{gasbudget}. The formation of this cold gas significantly enhances BHAR. It is important to note that the cold gas formation mechanism has been investigated in numerous previous works \citep{li14, li15, wang19, voit17, qiu19}, and our simulation's mechanism aligns with those findings. The formation of cold gas along the bubble boundary further supports the finding that the bubble displaces gas within the galaxy as it moves outward.

In the quiescent phase, hot gas mass within $5~R_{\rm e}$ is maintained at approximately $10^9~\mathrm{M_{\odot}}$. The total amount of hot gas is roughly consistent with observations within 5$R_{e}$ \citep{anderson13}. Meanwhile, the hot gas mass within $R_{\mathrm{vir}}$ gradually increases. In \citet{Zhu2023}, it is found that cosmological inflow does not enter the galaxy but instead becomes part of the circumgalactic medium. The growing hot gas mass within $R_{\mathrm{vir}}$ and the nearly constant hot gas mass within $5~R_{\rm e}$ align with this conclusion. In the quiescent state, cold gas within the galaxy is sparse, appearing only intermittently with peak cold gas mass values of $\la 10^{7}~\mathrm{M_{\odot}}$.  The low value of cold gas mass implies that our simulated galaxy is a single-phase galaxy (SPG), i.e., no cold gas can be detected. This is consistent with observations of \citet{young+14, voit2015nat}, which find that about half of the elliptical galaxies do not have cold gas. 

\subsection{The late stage evolution of massive elliptical galaxies}\label{explanation}

After the initial transition phase, the simulated galaxy remains in a quiescent state for the remainder of our simulation time. { The main goal of the present paper is to  investigate the role of AGN feedback in the late stage evolution of this quiescent state. In our model, the quiescent state of the galaxy is formed due to the strong outburst of the central AGN, which blows away a lot of gas out of the galaxy, as we have analyzed in the last section. On the one hand, this scenario is currently believed to be the mechanism of quenching a galaxy \citep[e.g.,][]{hopkins06}.  On the other hand, how the massive elliptical galaxy enters the quiescent state is not important for our aim, because the gas supply in the galaxy is dominated by the stellar wind from the evolution of old population while the initial gas setup is not so important \citep{Brighenti1999, mathews2003}. In fact, we have performed simulations with lower gas density as the initial condition thus there is no such AGN outburst. We find that the gas profile in the late-stage evolution almost remains unchanged compared to the case presented in this paper. }

\subsubsection{AGN feedback is required in the late stage evolution}
\label{overview}


\begin{figure}     
    \includegraphics[width=0.45\textwidth]{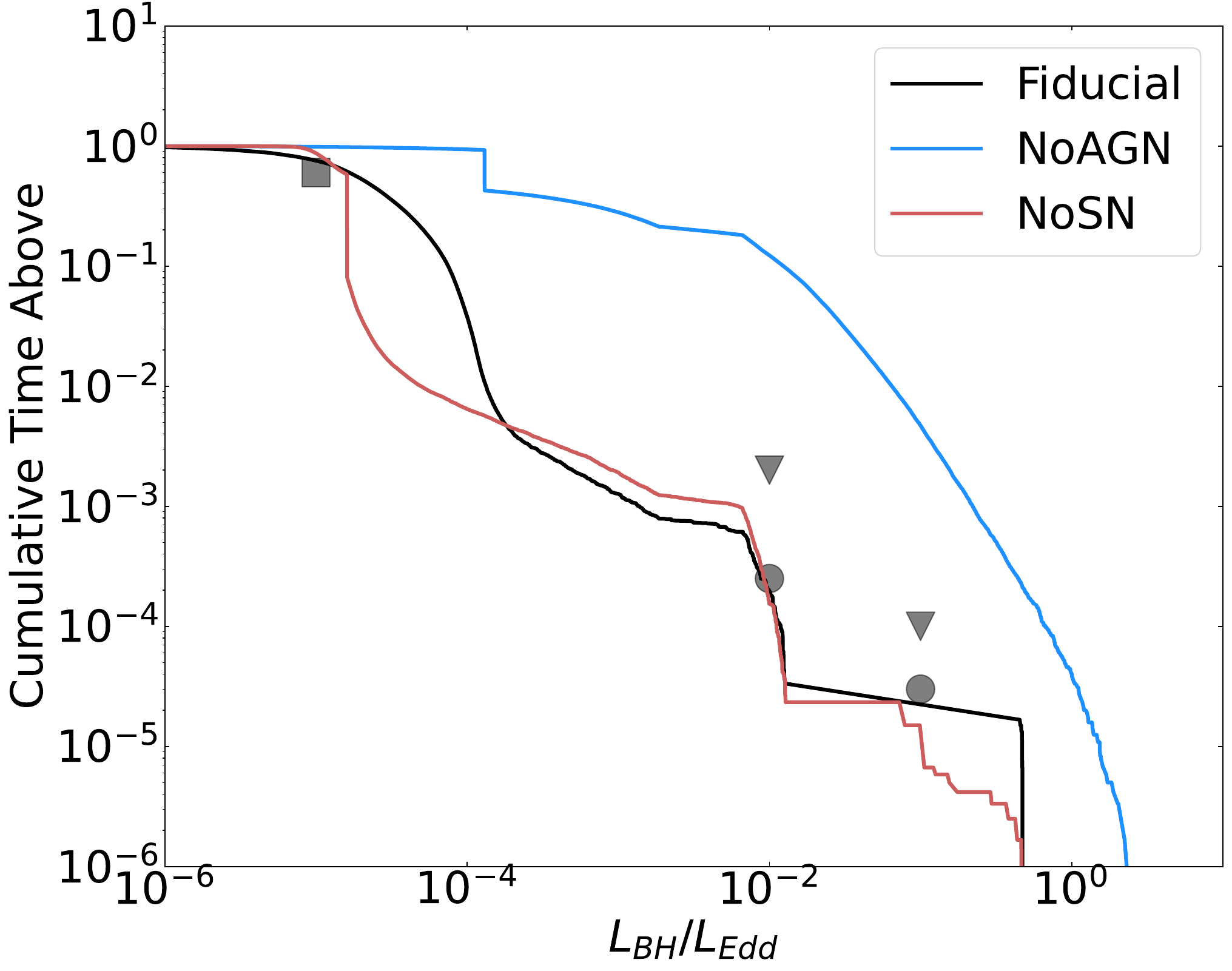}
    \caption{ The duty cycle of AGN predicted by ``Fiducial'', ``NoAGN'', and ``NoSN'' models. The squares, circles, and the  downward-pointing triangles represent the observational data from \citet{ho2009}, \citet{Greene07}, and  \citet{heckman2004}, respectively.}
    \label{agn_duty_cycle}
\end{figure}

To investigate this issue, we compared the BHAR and SFR { and the AGN duty cycle} of the ``Fiducial'', ``NoAGN'', and {``NoSN''} models. Recall that the latter two models are obtained by turning off the AGN and SN feedback from $t=6$ Gyr in the ``Fiducial'' model. 

Figure \ref{SFR&BHAR_total} displays the time evolution of the BHAR and SFR of these three models from 6 to 12 Gyr. In comparison to the ``Fiducial'' model, the ``NoAGN'' model exhibits a rapid increase in both SFR and BHAR during the initial $\sim$ 0.5 Gyr. The BHAR rises from $\sim 10^{-2}~\mathrm{M_{\odot}}$/yr to $\ga 1 ~\mathrm{M_{\odot}}$/yr. The latter value  corresponds to $\sim 10^{-1}\dot{M}_{\mathrm{Edd}}$, which is  over two orders of magnitude higher than the observed value of $\sim 10^{-3} \sim 10^{-4}\dot{M}_\mathrm{Edd}$ for black holes in quiescent massive galaxies \citep{ho08, ho2009}. { The BHAR in ``NoAGN'' model is about two orders of magnitude higher than the ``Fiducial'' model, reaching $\sim1{\rm M_{\odot}}$/yr. Such a high BHAR results in an unreasonably high black hole mass of nearly $10^{10}{\rm M_{\odot}}$.} The SFR increases by nearly three orders of magnitude, from $\sim 10^{-5}~\mathrm{M_{\odot}}$/yr to $\sim 10^{-2} ~\mathrm{M_{\odot}} $/yr. In contrast, the SFR value in the ``NoSN'' model increases only by a factor of $\sim$ 10 compared to the fiducial model, while the BHAR value even decreases by a factor of $\sim$ 2, possibly due to increased SFR consuming the gas in the galaxy.

{ Figure \ref{agn_duty_cycle} shows the duty-cycle of the AGN predicted by the ``Fiducial'', ``NoAGN'', and ``NoSN'' models. The simulation data is taken from 6 to 12 Gyr. From the figure, we can see that the AGN duty-cycle predicted by the ``Fiducial'' and ``NoSN'' models are very similar, while the duty-cycle predicted by the  ``NoAGN'' model is significantly different. This is consistent with the results shown in the left panel of Figure \ref{SFR&BHAR_total}. We further compare the duty-cycle in three models with observation. From the figure, we can see that the duty-cycles predicted by the ``Fiducial'' and ``NoSN'' models are consistent with observations, while that predicted by the ``NoAGN'' model is not.  

The above findings suggest that the AGN feedback will strongly suppress the growth of black hole and star formation, change  the AGN duty cycle, while the SN feedback has minor impacts on suppressing star formation and almost no effects on the black hole activities.} 
 
It is worth noting that the SFR value in the ``NoAGN'' model remains lower than the star-forming main sequence value, which is $\sim 10~\mathrm{M_{\odot}}$/yr for this galaxy mass. This is because our simulation focuses on low-angular momentum massive elliptical galaxies, whereas most star-forming galaxies are disk galaxies \citep{Lee2013}. In contrast to elliptical galaxies, stars in a disk galaxy have much higher specific angular momentum and gas concentration near the disk plane, resulting in a significantly higher star formation rate. As a test, we conducted a simulation of a high-angular-momentum elliptical galaxy and discovered that the SFR easily reached $10^{-1}~\mathrm{M_{\odot}}$/yr. We anticipate that if we were to simulate a disk galaxy without incorporating AGN feedback, the galaxy would reach the star-forming main sequence.

\subsubsection{Energetic analysis of various energy components}\label{energetics}

\begin{figure}     
    \includegraphics[width=0.45\textwidth]{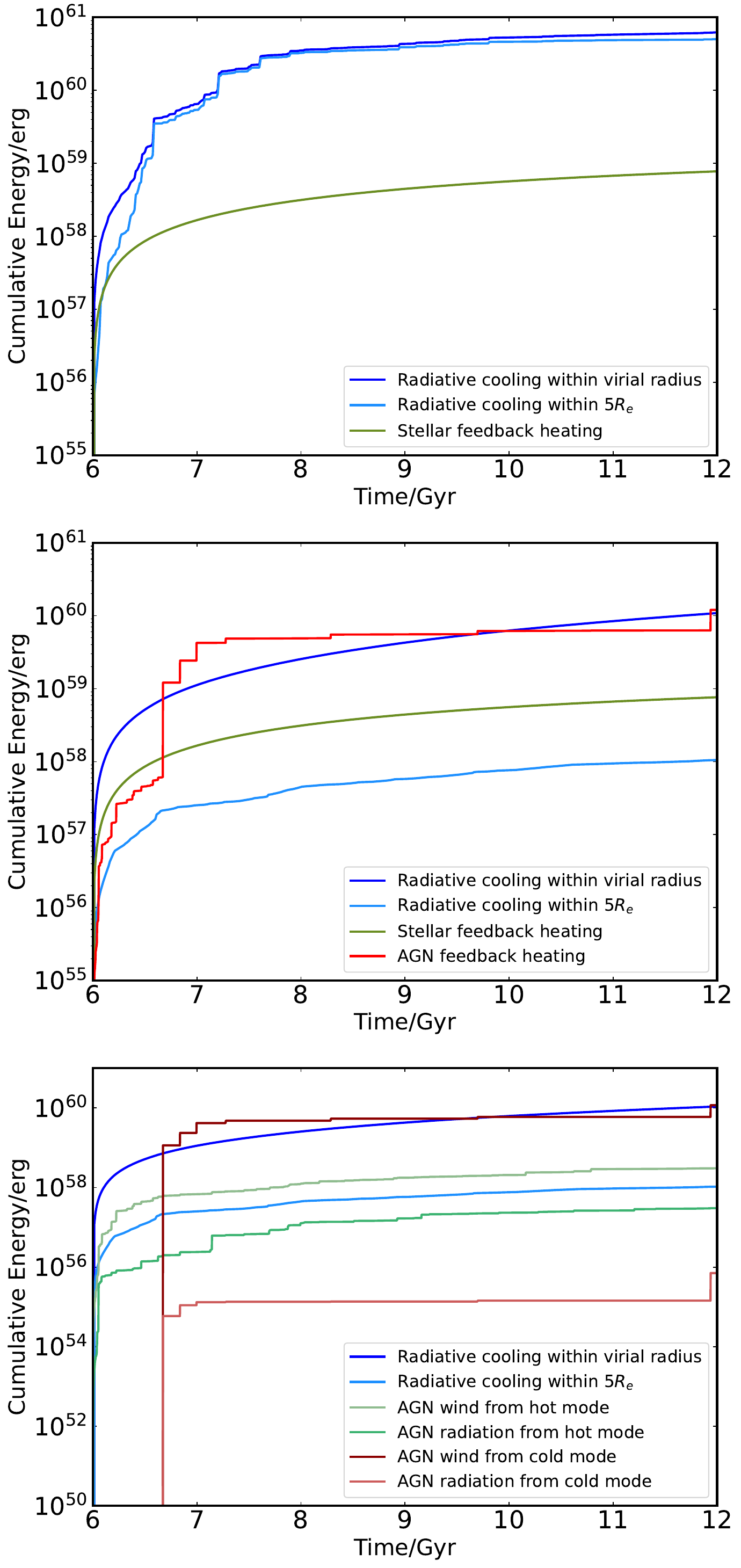}
    \caption{The cumulative energy of various components from 6 to 12 Gyr.  {\it Top:} Cumulative energy from stellar feedback heating in ``NoAGN'' model. {\it Middle:} Cumulative energy from AGN and stellar feedback heating in ``Fiducial'' model.  {\it Bottom:}  Cumulative energy from AGN wind and radiation from cold and hot modes  in ``Fiducial'' model. In each panel, the cumulative energy loss by radiative cooling within the virial radius and 5 $R_{\rm e}$ are shown. }
    \label{cooling&heating}
\end{figure}

Various energy components can compensate for the cooling of the gas in the galaxy, including those from AGN and stellar feedback. The former encompasses radiation and wind in both cold and hot modes\footnote{We have not considered jets in our model.}, while the latter includes heating from thermalized stellar wind and supernovae. In this section, we investigate the roles of these components from an energetic perspective.  We calculate radiative heating by considering the fraction of AGN radiation energy that has been absorbed by the { gas in the halo}. The AGN wind can deposit its energy into the { gas in the halo} through shock heating, PdV work by bubble expansion, dissipation of turbulence, sound waves, ``g-mode'' internal waves, and so on \citep{Churazov02, Schekochihin09, bambic19}, thus calculating wind energy absorption accurately is challenging.  On the other hand, according to \citet{Li17}, the energy dissipation of AGN jets in space may decrease rapidly along its propagation direction. So in the present work, we just simply assume that the AGN wind energy output is fully absorbed. This assumption is reasonable since most of the energy from the AGN wind will be deposited within $\sim 200$ kpc, as we will explain in section \ref{windimpact}. 

Figure \ref{cooling&heating} displays the cumulative values of various forms of energy from 6 to 12 Gyr in both ``NoAGN'' model and ``Fiducial'' model. In each panel, the cumulative energy by radiative cooling of the gas in the galaxy within $R_{\rm vir}$ and $5~R_{\rm e}$ are represented by deep and light blue lines, respectively. 
The top panel shows the results of the ``NoAGN'' model. We can see from this panel that the cumulative energy by radiative cooling within $R_{\rm vir}$ and $5~R_{\rm e}$ are almost same, indicating that the radiative cooling is dominated by the region within $5R_{\rm e}$. Additionally,  in this panel we illustrate the cumulative energy from the stellar outputs. We can see that it is several orders of magnitude lower than the  radiative cooling. This clearly indicates that stellar feedback alone is far from enough to compensate the radiative cooling of the gas in the galaxy.  

The middle and bottom panels show the results of the ``Fiducial'' model. We can see from these two panels that, different from the ``NoAGN'' model, the radiative cooling within $5R_{\rm e}$ in the ``Fiducial'' model is about two orders of magnitude lower than that within $R_{\rm vir}$. This is because the gas becomes much less concentrated toward the galaxy center due to the presence of AGN feedback. In the middle panel, we also illustrate the cumulative energy from both AGN and stellar outputs. From this panel, it becomes evident that the total energy from AGN output is comparable to the total energy loss from radiative cooling  within the whole halo, suggesting that AGN heating can compensate for radiative cooling. As a contrast, the energy output from stellar feedback is an order of magnitude lower than that of the radiative cooling within $R_{\rm vir}$, further emphasizing the significance of AGN feedback over stellar feedback in compensating for cooling loss, confirming our results in section \ref{overview} and previous research \citep{2018ApJ...866...70L,wang19}. On the other hand, different from the ``NoAGN'' model, we note that, in the ``Fiducial'' model, the stellar feedback heating is larger than the radiative cooling within $5~R_{\rm e}$. This implies that, with the help of AGN feedback, the stellar feedback outputs can  compensate the radiative cooling within $5R_{\rm e}$.

\begin{figure}     
    \includegraphics[width=0.45\textwidth]{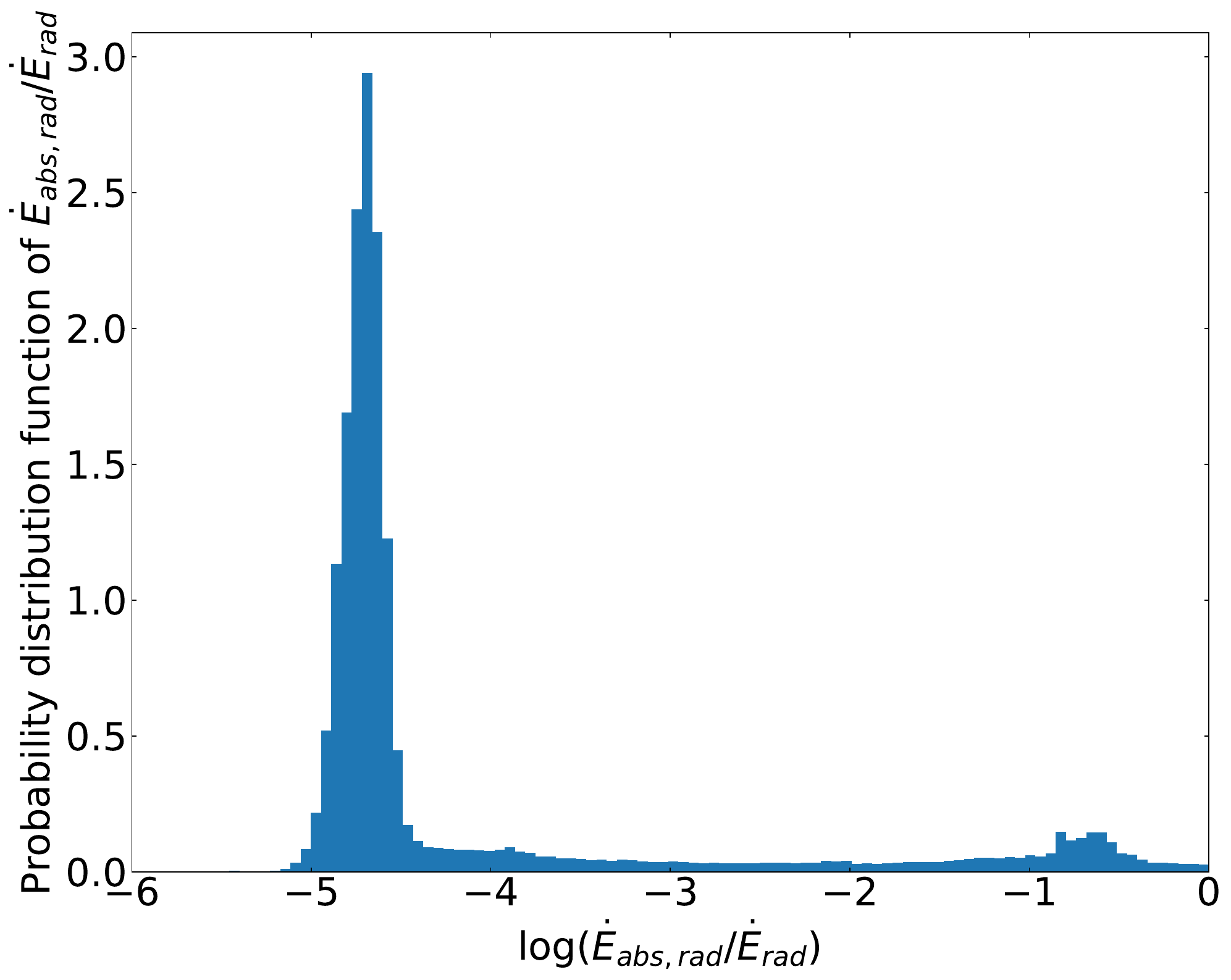}
    \caption{The probability distribution function of the ratio of the absorbed radiation power and the AGN luminosity.}
    \label{rad}
\end{figure}

In the bottom panel, in addition to the radiative cooling of the gas, we also present the cumulative energy output from wind and radiation for both hot and cold modes. From this panel, we observe that, in both hot and cold modes, radiative heating is less significant than wind heating. As the elliptical galaxy is predominantly composed of hot gas, its opacity primarily originates from Compton scattering, making it challenging for radiation to be absorbed and heat the gas. Figure \ref{rad} demonstrates the probability distribution of the ratio of the radiation power absorbed by { gas in the halo} and AGN radiation power at different times. The plot reveals that the ratio is mainly concentrated around $10^{-5}$, indicating that merely $\sim10^{-5}$ radiative energy from AGN can be absorbed and heat the gas. Due to the low absorption fraction, radiative energy is not particularly important, even in the cold mode, rendering wind energy output the dominant energy source in our simulations. However, we note that, in our model, we have not considered dust, which can substantially enhance the absorbed percentage of AGN radiation \citep{costa2018a, costa2018b, Barnes2020}. We plan to address the effect of dust in our future works.

Concerning wind heating, we can see from the middle panel of Figure \ref{cooling&heating} that the cumulative energy output from the cold-mode wind can compensate for the radiative cooling of the gas within the whole gaseous halo. This compensation provides a comprehensive view of the cooling and heating balance in the simulated galaxies. Figure \ref{cooling&heating} reveals that the cumulative energy from the hot-mode wind is about 30 times smaller and unable to compensate for the radiative cooling in the entire gaseous halo.  We have further examined the AGN duty-cycle in our simulation. Similar to the results shown in \citet{yuan18}, the ratio of the duration in the cold mode and the total AGN duration is around $10^{-4}$ in the final 6 Gyr, consistent with observations \citep[e.g.,][]{Greene07,ho2009}. This result indicates that, although the AGN spends most of its time in the hot accretion mode, its energy output is dominated by the short outburst of the cold mode. However, for the region within $5~R_{\rm e}$, we can see in the figure that the cumulative energy from the hot-mode wind is slightly higher than the radiative cooling. Consequently, in principle, the hot-mode wind can compensate for the radiative cooling within $5~R_{\rm e}$. Therefore, from the energetic argument alone, we cannot conclusively determine whether the hot-mode wind is unimportant. In the following section, we consider another constraint, i.e., investigating the spatial extent to which the wind in the two modes can heat { the gas in the halo}.

\subsubsection{The impact area of the AGN wind}
\label{windimpact}

The analysis in section \ref{energetics} suggests that, from an energetic standpoint, AGN wind heating is more crucial than AGN radiative heating, and the wind from the cold mode is likely more important than the wind in the hot mode. In this section, we further examine the spatial extent to which the wind from cold and hot modes can heat the hot gas halo. Although the heating mechanism of the AGN wind is complex, shock heating is dominant. Based on wind propagation theory, the wind will trigger a forward shock to heat the gas. The shock will slow down due to the pile-up of the { gas that the swept by wind}. When the shock speed is comparable to the sound speed of the { medium around the BH}, the shock will fade away and become a sound wave. Although the sound wave can still deposit energy into the {surrounded medium of the BH}, the proportion of energy in sound waves is approximately 30\% \citep{bambic19}. Therefore, we assume that the location where the shock fades away is the distance where the shock speed is comparable to the sound speed of { the gas at certain radius}. In the following, we estimate the value of this fade-away radius. 

\begin{figure}     
    \includegraphics[width=0.45\textwidth]{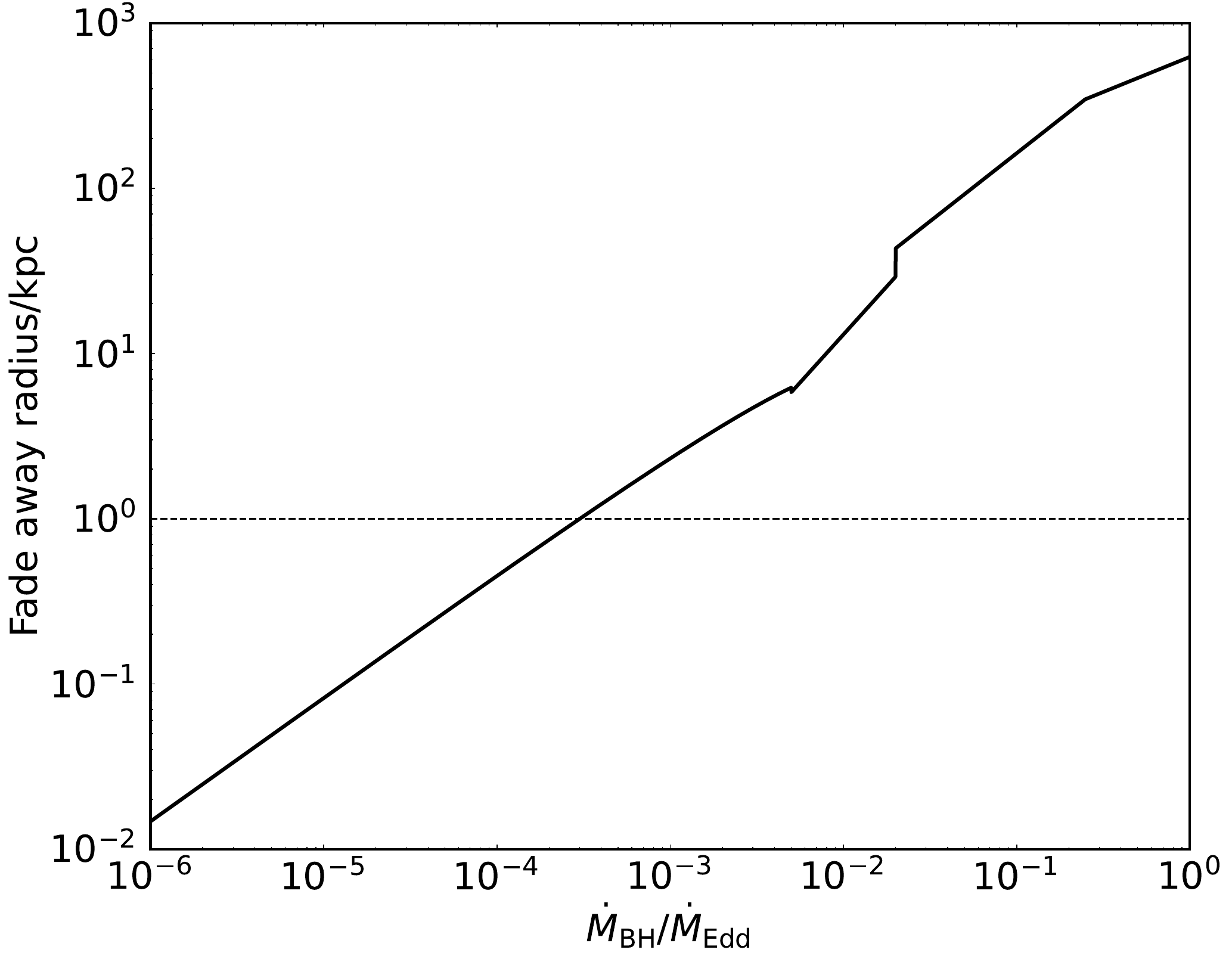}
    \caption{The fade away radius $R_{\rm fade}$ calculated through Equation \ref{fades} as a function of  black hole accretion rate.}
    \label{fade}
\end{figure}

\begin{figure}     
    \includegraphics[width=0.45\textwidth]{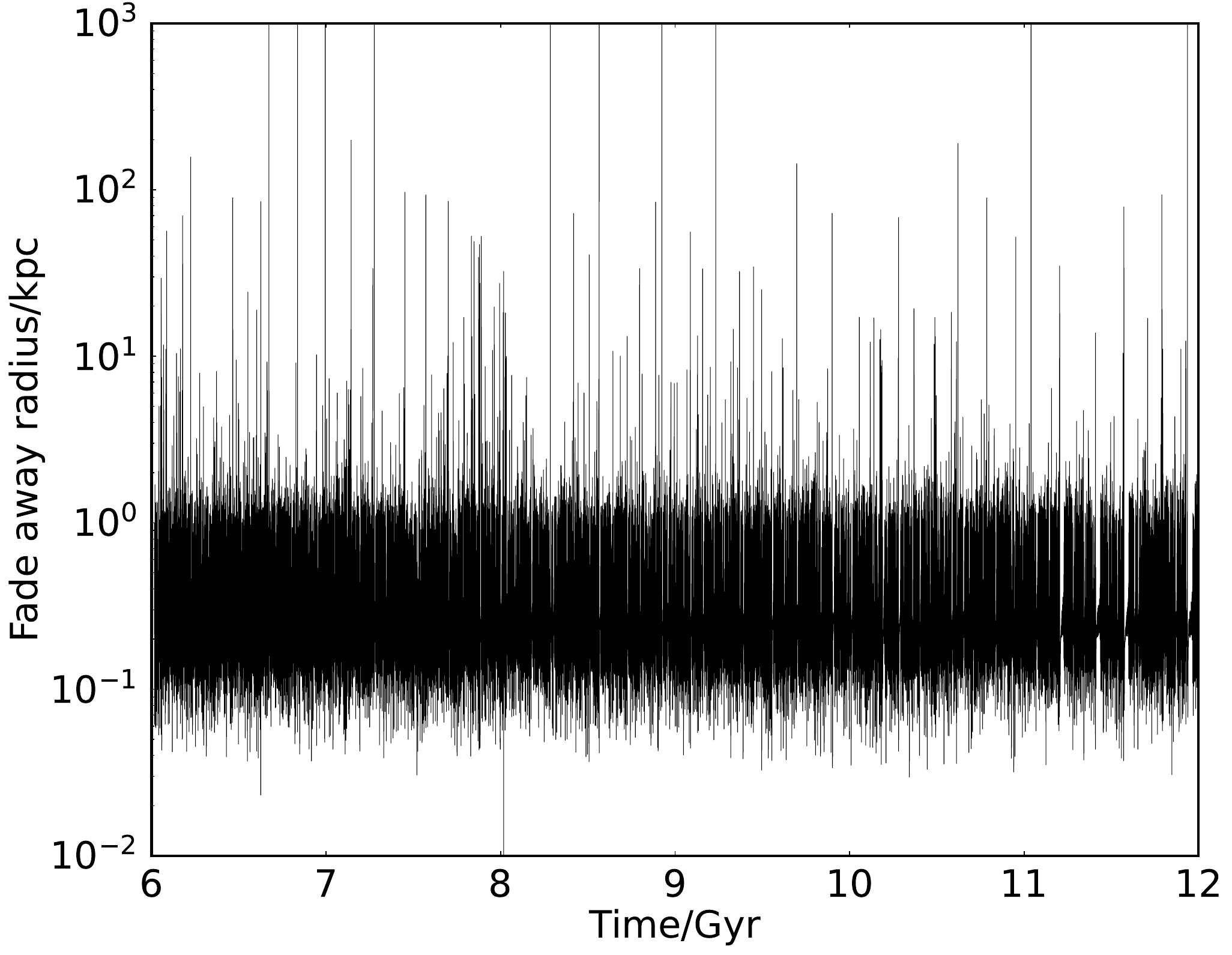}
    \caption{The time evolution of the fade away radius $R_{\rm fade}$ in our ``Fiducial'' model.}
    \label{fade_time}
\end{figure}

The { gas mass} accumulating in front of the shock depends on { the gas density around the BH}, denoted as $\rho_0$. Following the wind propagation theory \citep{ach2016}, we derive the location $R_{\rm s}$ and velocity $v_{\rm s}$ of the shock along its propagation direction as functions of the wind power $\dot{E}_{\mathrm{wind}}$:
\begin{equation}
R_{\rm s}\simeq \left(\frac{\dot{E}_{\mathrm{wind}}}{\rho_0}\right)^{1/5}t^{3/5},
\end{equation}
\begin{equation}
v_{\rm s}\simeq \frac{3}{5} \left(\frac{\dot{E}_{\mathrm{wind}}}{\rho_0}\right)^{1/5}t^{-2/5}.
\end{equation}
Here, $t$ denotes time. If the sound speed of { the gas around the BH} is $c_{\mathrm{s, gas}}$, the fade-away radius $R_{\rm fade}$ is estimated to be:
\begin{equation}\label{fades}
R_{\rm fade}\simeq \left(\frac{3}{5}\frac{(\dot{E}_{\mathrm{wind}}/\rho_0)^{1/5}}{c_{\mathrm{s,gas}}}\right)^{3/2} \left(\frac{\dot{E}_{\mathrm{wind}}}{\rho_0}\right)^{1/5}.
\end{equation}

As wind material propagates through { the medium around the BH}, a surface called the contact discontinuity (CD) forms behind the forward shock \citep{mou14}. The distance $d$ between the CD and the shock can be described by
\begin{equation}\label{distance}
d = \frac{\gamma-1}{3(\gamma+1)} R_s,
\end{equation}
where $\gamma$ is the adiabatic index of { the gas around the BH}.

In the above calculations, we have neglected radiative cooling, although it can cool the outer shell of the wind and affect the wind dynamics when the wind power is high \citep{king2015, Costa20}. The sound speed $c_{\mathrm{s, gas}}$ is assumed to be close to the stellar velocity dispersion, $\sigma_0=260~\mathrm{km\ s^{-1}}$. We also assume the density of the medium around the BH to be $\rho_0=1.67\times 10^{-25}~{\rm g\ cm^{-3}}$, which is comparable to the density of the gas located at approximately 100 pc in our simulations. Using these values, we can calculate the fade-away radius as a function of wind power.

Figure \ref{fade} displays the value of the fade-away radius as a function of AGN accretion rate $\dot{M}_{\rm BH}$ in units of Eddington accretion rate $\dot{M}_{\mathrm{Edd}}$ (approximately 50 $\mathrm{M_{\odot}\ yr^{-1}}$). Readers are referred to  Section \ref{AGNphysics} or \citet{yuan18} for the relation between $\dot{M}_{\rm BH}$ and the wind power $\dot{E}_{\rm wind}$. There is a slight discontinuity in our bridge formula for calculating the wind power at two accretion rates due to the transition between hot and cold accretion modes, which results in two discontinuities at $\dot{M}_{\mathrm{BH}}/\dot{M}_{\mathrm{Edd}}\simeq2\times 10^{-2}$ and $\dot{M}_{\mathrm{BH}}/\dot{M}_{\mathrm{Edd}}\simeq2\times 10^{-3}$. The fade-away radius at $\dot{M}_{\mathrm{BH}}/\dot{M}_{\mathrm{Edd}}=0.02$ is approximately 50 kpc, indicating that the wind in the cold mode can blow out of the galaxy and reach the circum-galactic region. The fade-away radius becomes smaller than 1 kpc when the black hole accretion rate is lower than approximately $3\times 10^{-4}\dot{M}_{\mathrm{Edd}}$. We have calculated the time evolution of the fade-away radius based on our simulation data from 6 Gyr to 12 Gyr; the results are shown in Figure \ref{fade_time}. In combination with Figure \ref{SFR&BHAR}, we can see that the fade-away radius is $\la 1$ kpc when the galaxy is in the quiescent state, i.e., the impact of the wind from the radio mode is limited to this region.

\begin{figure}     
    \includegraphics[width=0.45\textwidth]{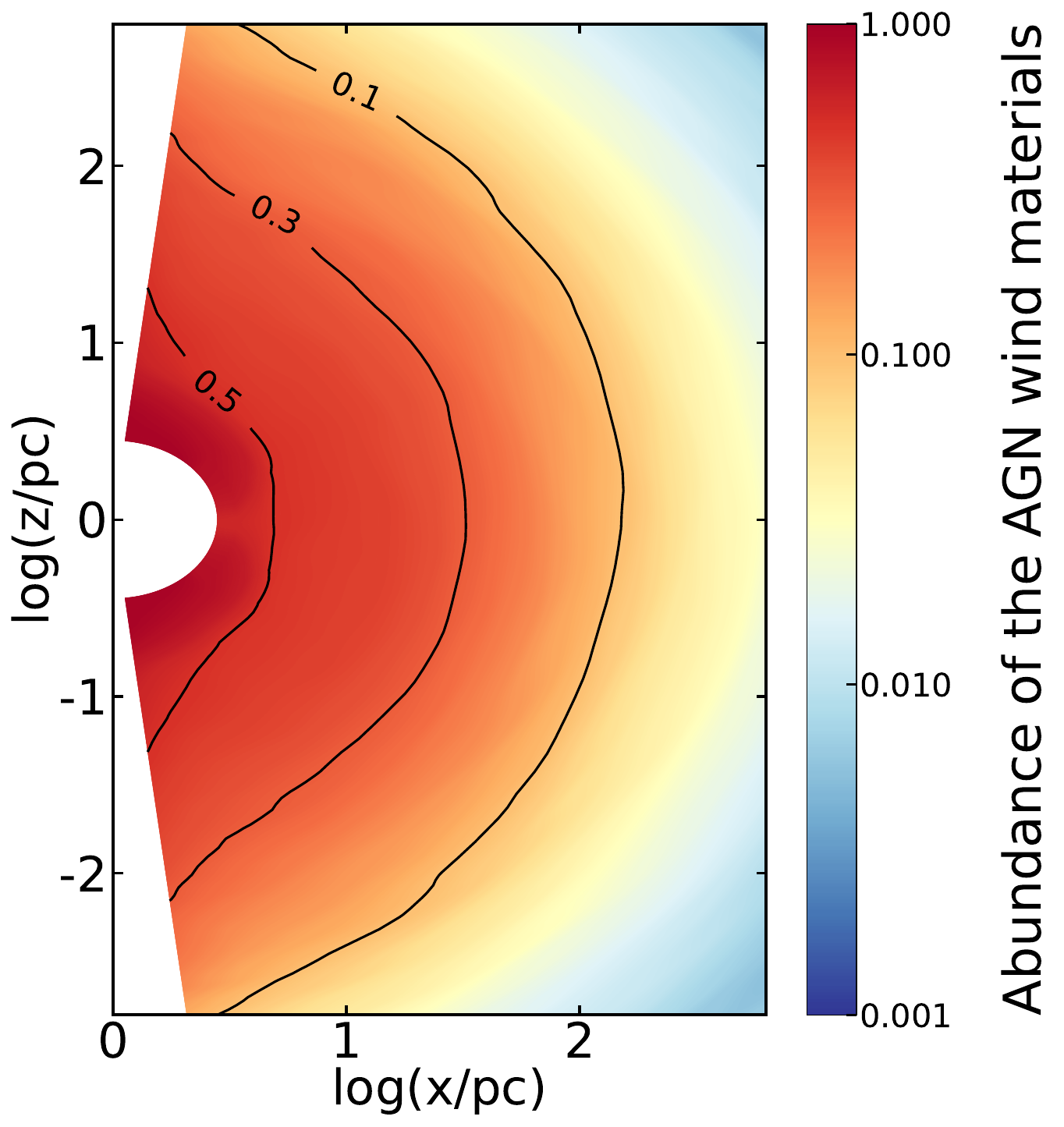}
    \caption{The spatial distribution of the averaged abundance of the materials of AGN wind, representing the proportion of the materials of the AGN wind in each gas cell. We take an average of 360 snapshots from 7.5 to 12 Gyr with 125 Myr time intervals. The black line represents the contours of the spatial distribution.}
    \label{abun}
\end{figure} 

\begin{figure}     
    \includegraphics[width=0.45\textwidth]{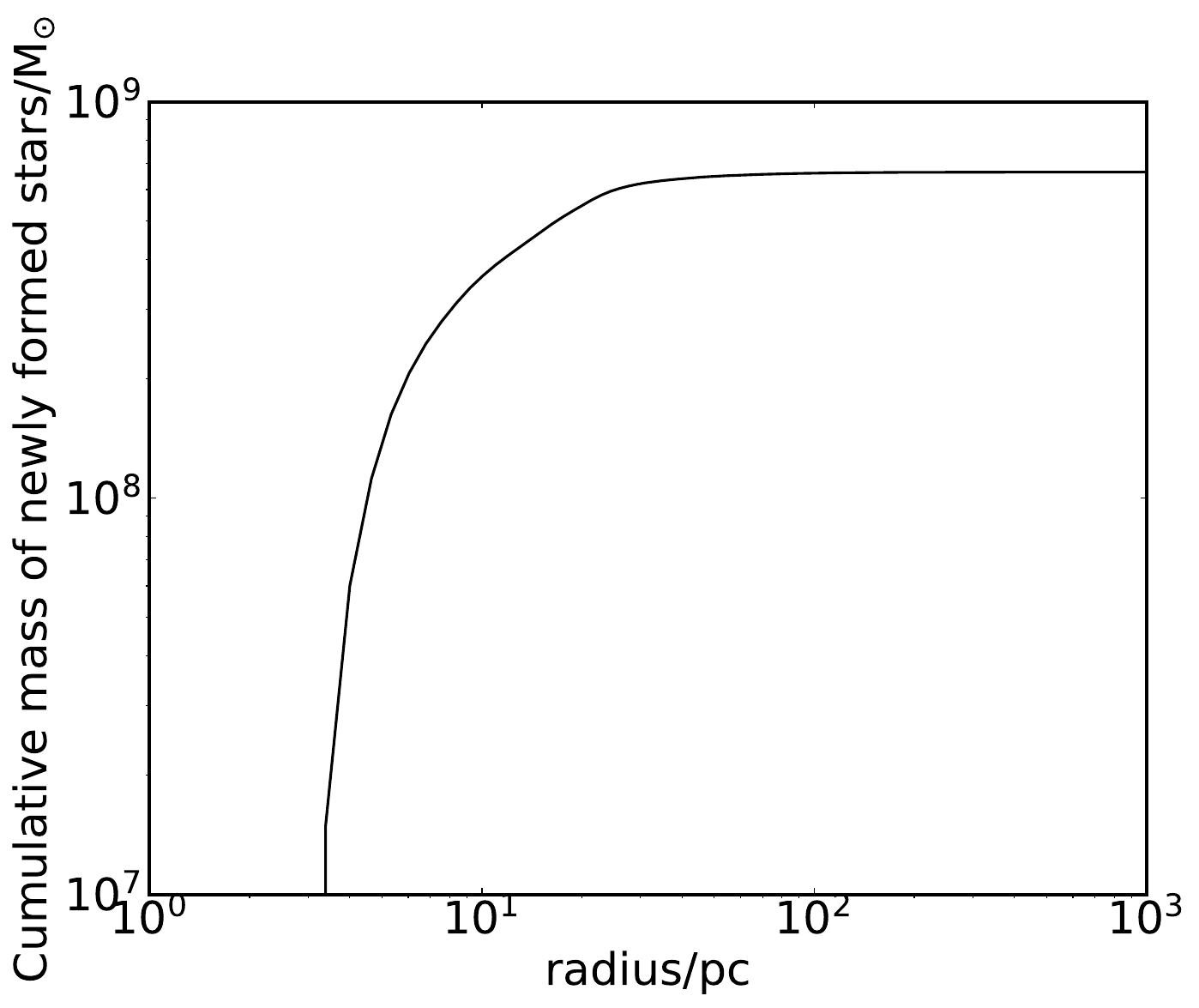}
    \caption{The radial distribution of the cumulative mass of newly formed stars from 7.5 Gyr to 12 Gyr.}
    \label{nfs}
\end{figure} 

We note that the fade-away radius presented in Figure \ref{fade_time} is a rough estimate based on certain approximations. Another way to investigate the actual propagation distance of the wind in a realistic environment is to trace the AGN wind materials in our simulation. Figure \ref{abun} displays the time-averaged spatial distribution of AGN wind material abundance, representing the proportion of AGN wind materials within a gas cell. The data is derived from the average of 360 snapshots taken between 7.5 and 12 Gyr at 125 Myr intervals.

Using Equation \ref{distance} with an adiabatic index of $\gamma=3/5$ for the ambient medium of the BH, we find that the distance $d$ between the contact discontinuity (CD) and the forward shock radius $R_s$ is $d=8.3\%R_s$, indicating that the CD is very close to the radius of the shock. In this case, wind materials cease near the fade-away radius. As the AGN wind's fade-away radius oscillates in simulations due to the AGN wind power and environmental variations, the time-averaged spatial distribution of AGN wind material abundance provides a good approximation of the spatial extent to which AGN wind can heat the {gas in the halo}.

The contour with a value of 0.1 is found to be approximately 300 pc, suggesting that the wind is unlikely to propagate beyond this radius, roughly consistent with our findings in Figure \ref{fade_time}. Additionally, Figure \ref{abun} reveals that the contour with a value of 0.3 is around 50 pc. Assuming that the wind impacts are significant when the abundance of AGN wind materials exceeds 30\%, this result implies that the wind has substantial effects only within 50 pc. As the black hole in the simulation remains in the hot mode for the majority of the time, these results hold primarily for the hot-mode wind.

Now we connect the above results with the region of star formation. Figure \ref{nfs} shows the radial distribution of cumulative mass of newly formed stars from 7.5 to 12 Gyr. From the figure, we can see that most of the stars are formed within 30 pc. This region is comparable to the ``impact region'' of hot-mode wind discussed above, which seemingly suggests that hot-mode wind can significantly affect star formation. However, this is not true. This is because there is a ``cooling radius'' at which the gas begins to cool significantly and fuel the star formation \citep{white1991, springel01, Croton06, Somerville2008, Guo2011}. To suppress star formation, the gas needs to be heated at the cooling radius \citep{Su2021}. However, the value of this cooling radius in our simulated galaxy is typically around 30 kpc, much larger than the ``impact region'' of hot-mode wind. This result implies that the hot mode wind can hardly affect the star formation.

\cite{yoon19} investigated the role of the hot feedback mode in AGN feedback by comparing the results of a ``test simulation'' and the fiducial model. They found that when they turned off the hot mode feedback (i.e., their ``nohotFB'' model), the cumulative mass of newly formed stars was enhanced only within approximately 4 pc. This finding is roughly consistent with our conclusion above, namely that the hot-mode wind can only suppress star formation well below around 30 pc.

\subsubsection{Sweeping up the gas originating from stellar mass loss}

\begin{figure}     
    \includegraphics[width=0.45\textwidth]{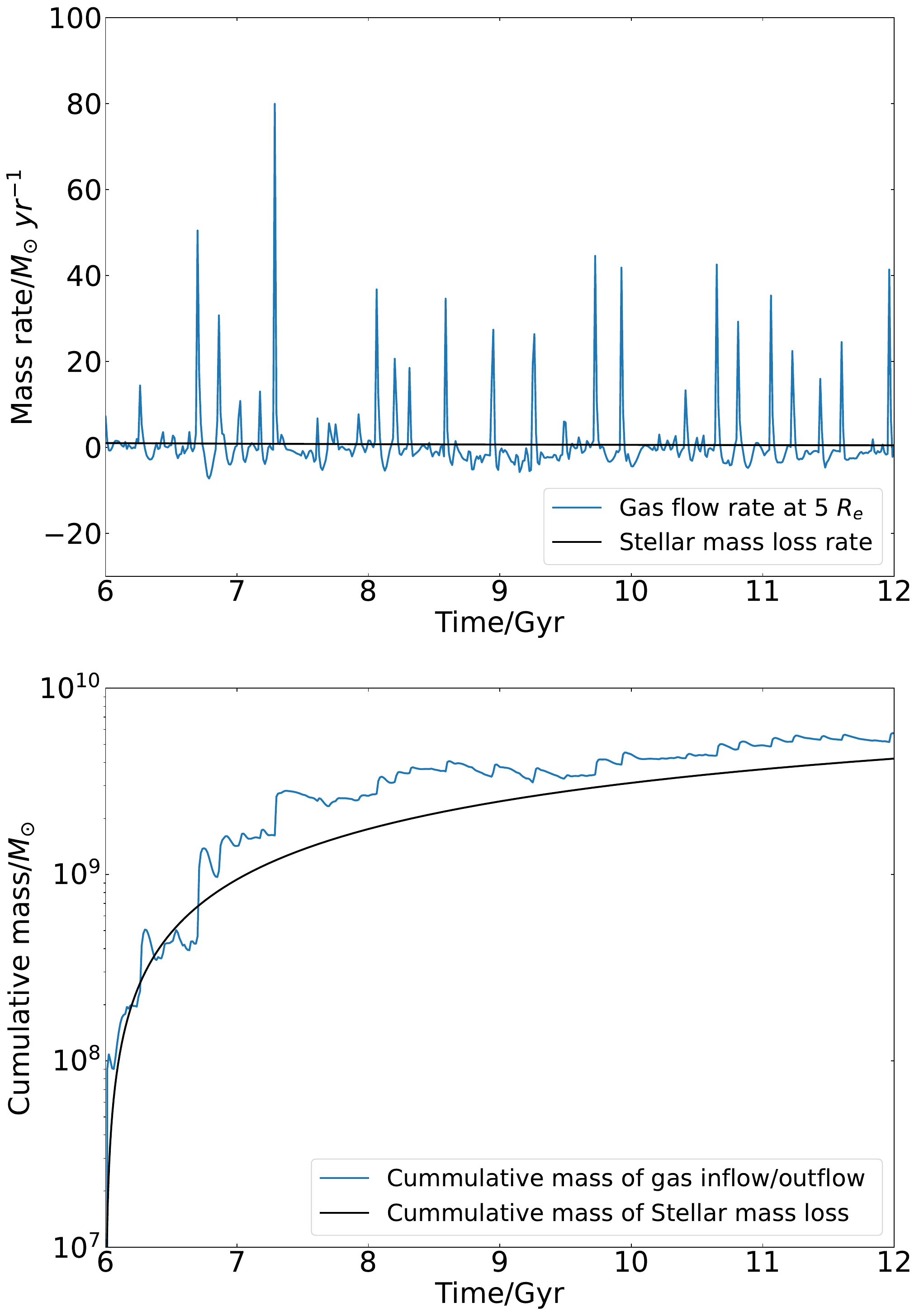}
    \caption{{\it Top panel:} The time evolution of gas flow rate at $5~R_{\rm e}$ (blue line) and stellar mass loss rate (black line). {\it Bottom panel:} The time evolution of the cumulative mass from gas inflow/outflow and from stellar mass loss from 6 to 12 Gyr. The gas flow rate is averaged with a 125 Myr time interval.}
    \label{sweep}
\end{figure} 

In order to maintain a quiescent state for both the AGN and the galaxy, it is necessary to sweep up the materials produced from stellar mass loss, as their cumulative mass can exceed $10^{10}~\mathrm{M_{\odot}}$ over the Hubble timescale. The black line in the top panel of Figure \ref{sweep} displays the time evolution of the stellar mass loss rate, which is approximately $1~\mathrm{M_{\odot}/yr}$, and the blue line displays the net mass rate (defined as the sum of inflow and outflow rates) at $5~R_{\rm e}$ from 6 to 12 Gyr. The black and blue lines in the bottom panel of Figure \ref{sweep} show the cumulative mass from stellar mass loss and the net rate within $5~R_{\rm e}$ as a function of evolution time. From the top panel, we observe that the gas flow rate at $5~R_{\rm e}$ is intermittent. From Figure \ref{snap_early}, we deduce that such intermittent behavior is driven by the AGN wind. More specifically, since { for the AGN wind in the hot mode, it is difficult to propagate out to 1 kpc} (section \ref{energetics}), we conclude that the intermittent outflow must be triggered by the AGN wind in the cold mode.

As shown by the bottom panel of Figure \ref{sweep}, the net flow of gas at $5~R_{\rm e}$ results in a generally positive cumulative mass of gas. We also observe that the value of this cumulative mass is similar to the cumulative mass from stellar mass loss, indicating that the total amount of gas escaping from $5~R_{\rm e}$ is roughly equal to the amount of gas produced from stellar mass loss. This clearly indicates that the intermittent outflow triggered by AGN feedback in the quasar/cold mode can efficiently sweep up the materials from stellar mass loss. This is yet another piece of evidence for the importance of cold mode AGN feedback.

\section{Summary and Discussions}\label{conclusion}

In this paper, we investigate the effect of AGN feedback on the late stage evolution of massive elliptical galaxies by performing numerical simulations of three models in the framework of {\it MACER}. In the ``Fiducial'' model, both AGN and supernovae feedback are included, and the galaxy consistently remains in a quiescent state. However, in the ``NoAGN'' and ``NoSN'' models, both the AGN feedback and supernovae feedback are turned off, respectively. These simulations are performed using the {\it MACER} framework \citep[e.g.,][]{yuan18}. The main advantages of {\it MACER} are that the simulation inner boundary is smaller than the Bondi radius of the black hole accretion flow, allowing for a reliable calculation of the mass accretion rate, and the adoption of state-of-the-art AGN physics at various accretion rates. By analyzing the simulation data of the ``Fiducial'' model and comparing its results with the ``NoAGN'' and ``NoSN'' models, we conclude that the wind launched by the AGN in the cold mode plays the dominant role in keeping the black hole at the center of the massive galaxy in low accretion state and suppressing the star formation. This conclusion is based on the following main findings:

\begin{enumerate}[1.]

\item We have compared the SFR and BHAR in the ``Fiducial'' and ``NoAGN'' models (section \ref{overview}). We find that the BHAR increase by around two orders of magnitude when AGN feedback is turned off, from $\sim 10^{-2} ~ \mathrm{M_{\odot}}$/yr ($\sim 10^{-4} ~ \dot{M}_{\mathrm{Edd}}$) to $\sim 1 ~ \mathrm{M_{\odot}}$/yr ($\sim 10^{-2} ~ \dot{M}_{\mathrm{Edd}}$). As a comparison, the value of BHAR in the ``NoSN'' model is similar to the ``Fiducial'' model. The value of BHAR in the ``NoAGN'' model is systematically higher than the observations. While the SFR increases by around three orders of magnitude when AGN feedback is turned off, from $\sim 10^{-5}~\mathrm{M_{\odot}}$/yr to $\sim 10^{-2}~\mathrm{M_{\odot}}$/yr. These results indicate that AGN feedback is required in the late time stage.

\item There are various sources of feedback energy, including wind and radiation from the AGN and stellar feedback. We have calculated the cumulative energy of these components that are absorbed by the gas in the galaxy and compared it with the radiative cooling of the gas in the galaxy. We find that the AGN feedback energy can compensate for the radiative cooling within the virial radius, while SN feedback alone cannot (top and middle panels of Figure \ref{cooling&heating}). Moreover,  although the AGN spends most of its time in the hot mode in our simulation, which is consistent with observations,  the feedback energy from cold-mode wind is the dominant component, much larger than those from the hot-mode wind and radiation (bottom panel of Figure \ref{cooling&heating}), and can compensate for the radiative cooling of the whole halo. However, the energy from the hot-mode wind can also compensate for the radiative cooling within $5~R_{\rm e}$. This implies that energetic arguments cannot rule out the possibility that hot-mode wind can keep the galaxy quiescent.

\item We have estimated the spatial extent to which the AGN wind can heat the { gas in the halo}. Such heating is mainly via the shock produced by the wind-medium interaction. A rough estimation, characterized by the ``fade-away radius'', indicates that the extent should be smaller than 1 kpc (Figure \ref{fade_time}). A more accurate estimation obtained by the time-averaged abundance of wind material (Figure \ref{abun}) shows that most of the time (i.e., when the AGN is in the hot mode), the wind can only heat the { gas around the BH} and affect star formation up to $\sim50$ pc. This value is much smaller than the ``cooling radius'' at which strong heating must occur to prevent the gas from cooling and forming stars. This implies that the hot-mode wind is not efficient in suppressing star formation and keeping the galaxy in a quiescent state. 

\item To keep the galaxy and black hole in a quiescent state, it is also necessary to sweep up the gas produced by stellar mass loss. We have calculated the gas flow rate at $5~R_{\rm e}$ as a function of time and found some strong outflow spikes (top panel of Figure \ref{sweep}), which should be attributed to the strong cold-mode AGN wind. We have also calculated the cumulative mass due to the gas inflow/outflow at $5~R_{\rm e}$ and found that this mass is consistent with the cumulative mass from stellar mass loss (bottom panel of Figure \ref{sweep}). This result implies that the strong cold-mode wind can efficiently sweep up the material from the stellar mass loss.

\end{enumerate}

 Although the hot-mode feedback does not play the dominant role in suppressing star formation, it still plays an important role in controlling the black hole growth and AGN luminosity, as pointed out by \citet{yoon19}. Their simulations indicate that, if we were to turn off the hot-mode feedback, the growth of black hole mass $\Delta M_{\rm BH}$ during 12 Gyr evolution would increase from $\sim 0.05\times 10^9 ~\mathrm{M_{\odot}}$ to $\sim 0.7 \times 10^9 ~\mathrm{M_{\odot}}$. In addition, the typical hot-mode AGN luminosity would become about two orders of magnitude higher.

One caveat of our model is that we have not considered the jet in hot feedback mode, which should always exist \citep{yuan14}. A natural question then arises: whether the inclusion of jets change our conclusion? Using three-dimensional general relativity MHD simulations of hot accretion flows around black holes, \citet{yang21} investigated the properties of jets and winds. They considered both SANE (standard and normal evolution) and MAD (magnetically arrested disk) and various black hole spins. They found that the power of a jet can be 10 times higher than that of a wind, depending on the accretion mode and black hole spin; while the momentum flux of the wind is always somewhat larger than that of the jet. It is worth noting that, in this paper, we argue against the hot-mode wind not because its power is insufficient, but because it cannot heat gas at large distances. Although the momentum flux of a jet is smaller than that of a wind, the opening angle of a jet is about an order of magnitude smaller than that of a wind. Therefore, the jet can pierce the gaseous halo and deposit its energy at much larger and more ``suitable'' distances than the hot-mode wind to suppress cooling in the halo. In this context, \citet{Su2021} studied the role of jets in galaxy evolution and found that they can maintain the quiescent state of the galaxy. Then an interesting question is that, among jet and cold-mode wind, which one plays the dominant role or do they play a comparable role in maintaining the quiescent state of the galaxy. We will try to answer this question in our future work. The gas fraction obtained in our simulations is about $\sim 20\%$, which is somewhat lower than the observed value of $30-40\%$ according to some definition of the gas fraction and radius \citep{mccarthy2017}. If the gas fraction in our simulations were higher,  the role of halo gas cooling would be more important, and it would be more difficult to maintain these galaxies in a quenched state. 

\section*{Acknowledgments}
{ We thank the anonymous referee for his/her detailed comments.} BZ, FY, and SJ are supported in part by the NSF of China (grants 12133008, 12192220, and 12192223), YP by the NSF of China (grants  12125301, 12192220, and 12192222) and the science research grants from the China Manned Space Project (CMS-CSST-2021-A07), LCH by the NSF of China (grants 11721303, 11991052, 12011540375, and 12233001) and the China Manned Space Project (CMS-CSST-2021-A04, CMS-CSST-2021-A06).  The calculations have made use of the High Performance Computing Resource in the Core Facility for Advanced Research Computing at Shanghai Astronomical Observatory.

\section*{Data Availability}

The data underlying this article will be shared on reasonable request to the corresponding author.



\bibliographystyle{mnras}
\bibliography{ref} 



\bsp	
\label{lastpage}
\end{document}